\documentclass{elsart}

\usepackage[square,comma]{natbib}
\usepackage{graphicx}
\usepackage{pxfonts}
\usepackage{lineno}

\usepackage{amssymb}

\journal{}

\begin{document}

\thispagestyle{empty}
\begin{Large}
\textbf{DEUTSCHES ELEKTRONEN-SYNCHROTRON}

\textbf{\large{Ein Forschungszentrum der Helmholtz-Gemeinschaft}\\}
\end{Large}

DESY 11-134

August 2011

\begin{eqnarray}
\nonumber &&\cr \nonumber && \cr \nonumber &&\cr
\end{eqnarray}
\begin{eqnarray}
\nonumber
\end{eqnarray}
\begin{center}
\begin{Large}
\textbf{Scheme for generating and transporting THz radiation to the
X-ray experimental floor at the LCLS baseline}
\end{Large}
\begin{eqnarray}
\nonumber &&\cr \nonumber && \cr
\end{eqnarray}

\begin{large}
Gianluca Geloni,
\end{large}
\textsl{\\European XFEL GmbH, Hamburg}
\begin{large}

Vitali Kocharyan and Evgeni Saldin
\end{large}
\textsl{\\Deutsches Elektronen-Synchrotron DESY, Hamburg}
\begin{eqnarray}
\nonumber
\end{eqnarray}
\begin{eqnarray}
\nonumber
\end{eqnarray}
ISSN 0418-9833
\begin{eqnarray}
\nonumber
\end{eqnarray}
\begin{large}
\textbf{NOTKESTRASSE 85 - 22607 HAMBURG}
\end{large}
\end{center}
\clearpage
\newpage

\begin{frontmatter}



\title{Scheme for generating and transporting THz radiation to the X-ray experimental floor at the LCLS baseline}


\author[XFEL]{Gianluca Geloni\thanksref{corr},}
\thanks[corr]{Corresponding Author. E-mail address: gianluca.geloni@xfel.eu}
\author[DESY]{Vitali Kocharyan}
\author[DESY]{and Evgeni Saldin}

\address[XFEL]{European XFEL GmbH, Hamburg, Germany}
\address[DESY]{Deutsches Elektronen-Synchrotron (DESY), Hamburg,
Germany}

\begin{abstract}
This paper describes a novel scheme for integrating a coherent THz
source in the baseline of the LCLS facility. Any method relying on
the spent electron beam downstream of the baseline undulator should
provide a way of transporting the radiation up to the experimental
floor. Here we propose to use the dump area access maze. In this way
the THz output must propagate with limited size at least for one
hundred meters in a maze, following many turns, to reach the near
experimental hall. The use of a standard, discrete, open
beam-waveguide formed by periodic reflectors, that is a mirror
guide, would lead to unacceptable size of the system. To avoid these
problems, in this paper we propose an alternative approach based on
periodically spaced metallic screens with holes. This quasi-optical
transmission line is referred to as an iris line. We present
complete calculations for the iris line using both analytical and
numerical methods, which we find in good agreement. We present a
design of a THz edge radiation source based on the use of an iris
line. The proposed setup takes almost no cost nor time to be
implemented at the LCLS baseline, and can be used at other
facilities as well. The edge radiation source is limited in
maximally achievable field strength at the sample. An extension
based on the use of an undulator in the presence of the iris line,
which is feasible at the LCLS energies, is proposed as a possible
upgrade of the baseline THz source.
\end{abstract}

%
%
%
\end{frontmatter}



\section{\label{sec:uno}  Introduction}

The accelerator complex at the LCLS produces ultra-short electron
bunches approaching sub-hundred fs duration. It is natural to take
advantage of these ultra-short bunches in order to provide coherent
THz radiation \cite{CDRL2}. In fact, intense, coherent THz radiation
pulses can be produced from the sub-hundred fs electron bunches at
wavelength longer than, or comparable with the bunch length, leading
to radiated energy levels proportional to the square of the electron
number, in contrast to the incoherent case when energy in the
radiation pulse scales linearly with the number of electrons
involved in the process. The result is an enhancement in radiation
intensity of up to $9-10$ orders of magnitude.

The exploitation of such kind of coherent THz source as a part of
the LCLS-II baseline user facility has been proposed in the LCLS-II
Conceptual design report \cite{CDRL2}.  THz radiation pulses can be
generated by the spent electron beam downstream of the X-ray
undulator. In this way, intrinsic synchronization with the X-ray
pulses can be achieved. A first, natural application of this kind of
the photon beams is for pump-probe experiments. Through the
combination of THz pump and X-ray probe, LCLS-II would offer unique
opportunities for studies of ultrafast surface chemistry and
catalysis \cite{CDRL2}. Also, the LCLS team started an $R\&D$
project on THz radiation from the spent electron beam downstream of
the undulator \cite{GALA}. The THz is generated by inserting a thin
Be foil into the electron beam \cite{ZWUP}. In this paper we
describe a novel scheme for integrating such kind of source in the
baseline of the LCLS facility.

The transport of the THz radiation from the LCLS beam dump area to
the near experimental hall constitutes a challenge. A major
constraint for the LCLS baseline case is constituted by the upstream
shielding wall and by the Far End Enclosure (FEE) downbeam shielding
wall. It follows that transport of the THz beam can only be achieved
using the dump area access maze, and relying on a limited size
\cite{MOEL,SANT}. However, in this way, the THz output must
propagate at least $100$ meters to reach the near experimental hall.
Since THz beams are prone to significant diffraction, a suitable
beam transport system must be provided to guide the beam along large
distances maintaining it, at the same time, within a reasonable
size. Moreover, the THz beamline should be designed to obtain a
large transmission efficiency for radiation over a wide wavelength
range. Usually, the focusing of the THz beam can only be achieved
with reflective optics because lenses made from any material would
reflect and absorb all radiation. In order to cope with the
unacceptable size increase, in this paper we propose an alternative
solution to a mirror guide, based on the use of periodically spaced
metallic screens with holes. This quasi-optical transmission line is
referred to as an iris line, or an iris beam waveguide. The
eigenmodes of the iris line have been calculated numerically for the
first time by Fox and Li \cite{FOX1} and later obtained analytically
by Vainstein \cite{VAI1,VAI2}.  When the Fresnel number of the iris
line is large, eigenmodes are characterized by rather small
diffraction losses. For instance, at a wavelength $\lambda \sim 0.1$
mm, for an iris radius $a \sim 5$ cm, and for a the distance between
the iris $b \sim 30$ cm, the Fresnel number is given by $a^2/(
\lambda b) \sim 10^2$, and diffraction losses of the principal
eigenmode are found to be about $10 \%$ in $100$ meters. Also, the
iris line is rather stable with respect to screens misalignments.
When the screens are adjusted in transverse and longitudinal
directions with an accuracy better than $1$ mm, misalignment do not
result in extra diffraction losses within the THz wavelength range.

In this paper we present complete iris line theory calculations. In
particular, the iris line eigenmodes are studied. The analysis of
the transmission line has been performed in two steps: the first one
consists of numerical simulations employing the method by Fox and Li
\cite{FOX1}, the second one by analytical calculations using the
method by Vainstein \cite{VAI1,VAI2}. Numerical simulation results
are in good agreement with analytical results.

In order to efficiently couple radiation into the transmission line,
it is desirable to match the spatial pattern of the source radiation
to the mode of the transmission line. To this end, it is advisable
to generate radiation from the spent electron beam directly in an
iris line with the same parameters $a$ and $b$  used in transmission
line. In this way, the source generates THz radiation pulses with a
transverse mode that automatically matches the mode of the
transmission line. We developed a theory supporting this choice of
THz source. As for the microwave waveguide case, one can use a
Green's function approach to solve the field equations. According to
method by Vainstein \cite{VAI1}, one can set complex boundary
conditions for the field on the virtual side surface of the iris
line, which are called impedance boundary conditions. The problem of
an open waveguide excitation is thus reduced to that of a closed
waveguide. Our consideration is quite general, and can be applied to
edge radiation sources as well as to undulator radiation sources in
the presence of an iris line.

We present a complete design for a THz edge radiation source at the
LCLS baseline. It includes a $15$ m-long electron beam vacuum
chamber equipped with an iris line, and a $100$ m-long transmission
line with the same parameters. The transmission line, which develops
through the access maze, presents six $90$ degrees turns with plane
mirrors at $45$ degrees as functional components. It is possible to
match incident and outgoing radiation without extra losses in these
irregularities. The proposed setup takes almost no cost nor time to
be implemented at the LCLS baseline.

The THz edge radiation source is limited concerning the achievable
field strength. An optimal expansion strategy for the LCLS THz
source would include a THz undulator source. Such modern, high power
THz source is exemplified by devices such as the coherent THz
undulator source at FLASH at DESY \cite{FAAT,GENS}. In this paper we
will describe this likely extension and its accommodation together
with the proposed edge radiation source. Both options have
advantages, and disadvantages. The edge radiation source is
characterized by low cost and can operate at all electron beam
energies, but is limited in field strength. The THz undulator source
would provide increase in field strength, but a trade-off must be
reached between the far-infrared and the X-ray achievable frequency
range due to technical limitations on magnetic field strength of
room-temperature electromagnetic undulators\footnote{On the one
hand, a decreased electron energy extends the far-infrared frequency
range. On the other hand, it limits the X-ray frequency range.}.

\section{\label{sec:due}  Principles of THz radiation generation based on the use of an iris line}

\subsection{Coherent THz edge radiation source}

As discussed above, the availability of a THz source at XFEL
facilities should be complemented by the availability of a suitable
THz beam transport system, which must guide the beam for distances
in the $100$ meters range.

\begin{figure}[tb]
\includegraphics[width=1.0\textwidth]{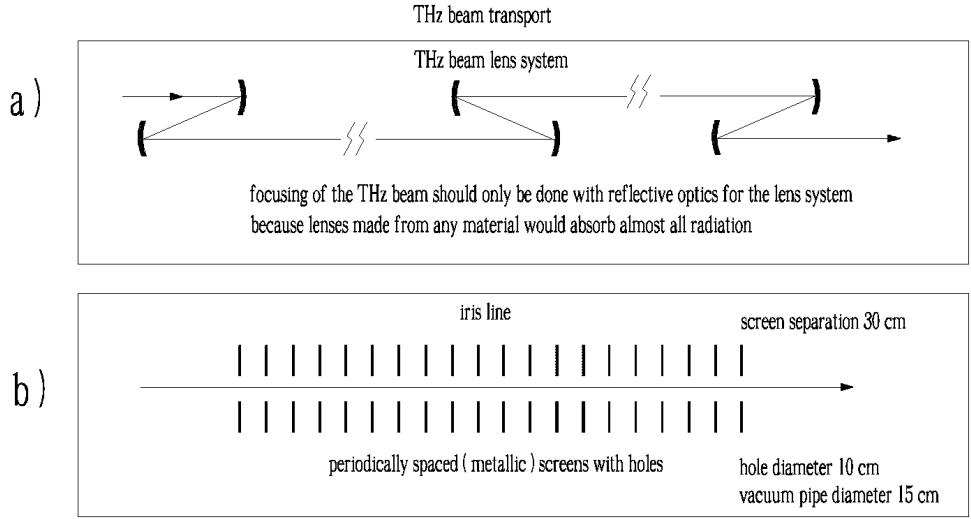}
\caption{Optical arrangement for the THz beam transport system. (a)
THz beam lens system. Focusing of the THz beam can only be performed
with reflective optics. In fact, lenses made of any material (except
diamond) would absorb almost all radiation. (b) The use of an iris
line, made of periodically spaced metallic screens with holes is a
way to obtain a high transfer efficiency at small diameter of the
beam pipe.} \label{th1}
\end{figure}
The THz beam can only be transmitted with quasi-optical techniques.
In particular, the idea of providing a periodic phase correction for
the free-space beam, in order to compensate for its divergence, is
very natural. In the 1960s numerous attempts of designing various
quasi-optical transmission lines were reported. In particular, it
was proposed to use open beam waveguides such as lens guides, mirror
guides, and iris guides \cite{FOX1}-\cite{FERN}. The competition
among different proposals ended with the victory of mirror guides,
still in use today e.g. for plasma heating \cite{SORO,FERN}.

Focusing of the THz beam can only be provided with reflective optics
since lenses made of any material would reflect and absorb all
radiation at long distances\footnote{In additional to absorption
losses, a dielectric lens suffers from reflections at its boundaries
as well. These surface reflections include a number of effects that
are almost always harmful.} . Fig. \ref{th1} (a) shows the optical
arrangement of a transport system based on the use of a mirror
guide. Each focusing unit is composed by two matched copper mirrors.
The mirrors are separated in such a way that the incident angle is
sufficiently small to minimize astigmatism. Existing mirror guides
are characterized by a maximal length of about $40-60$ m
\cite{KULI,TIED}. However, at XFEL facilities, scientists need a
beam transport system working for significantly longer distances.
The transverse size of the line is critical for this kind of
applications of an open waveguide. Since at LCLS baseline the open
guide should be installed within a maze with limiting size,
advantages of the mirror guide are not evident.

In order to keep the transverse size of the guide to an acceptable
level we propose, as an alternative to the mirror line solution, to
use the iris line in Fig. \ref{th1} (b). Iris lines are
characterized by a low attenuation of the fundamental mode,
self-filtering of higher order modes, wide operational wavelength
range and mechanical integrity of the structure. The first
investigation of the influence of diffraction effects on the
formation of the field eigenmodes in an iris line was carried out by
Fox and Li using physical optics techniques \cite{FOX1}. According
to their explanation, an iris line operates in the following way.
Consider an electromagnetic wave passing inside the sequence of
iris. When the wave diffracts at the first iris it produces a
diffraction pattern in the plane of the next iris. If the second
iris lets the main maximum of the diffraction pattern through,
diffraction losses are minimal. Further on, sideband maxima of the
diffraction pattern produced by the second iris are smaller than
those in the first pattern. After the wave has passed a large number
of irises, a field eigenmode is formed which has low diffraction
losses.

A very different and mathematically solid approach to the same
problem was introduced by Vainstein \cite{VAI1,VAI2}. His studies
were based on direct solution of Maxwell equations. They showed that
the way the electromagnetic field is confined inside an iris line,
is essentially different from the guiding mechanism in stable lens
guides or mirror guides. In fact, for iris guides, both diffraction
and reflection from the iris edges are involved. An analysis of this
effect allowed Vainstein to derive for the first time analytic
expressions for field distribution and mode losses in iris guides.
Complex boundary conditions are set on the virtual side surface of
the iris line, the so called impedance boundary conditions, and the
problem of an open waveguide transmission is therefore reduced to
that of a closed waveguide. An iris line has distinct transverse
modes. These eigenmodes are similar albeit not identical to
microwave waveguide modes. For instance, one can find that iris line
modes, in contrast with microwave waveguide modes, are independent
of the polarization of the radiation.

An infinite discrete set of eigenfunctions with corresponding
complex eigenvalues can be found. This set of modes comprises leaky
modes which are non-orthogonal and non-normalizable with respect to
the usual definition of inner product. To overcome this difficulty
one can define a bilinear form which, for our purpose, is equivalent
to an inner product under which modes are orthonormal.
Bi-orthogonality is often exploited in different problems
\cite{STRO}. For example, it can be shown that the treatment of a
microwave overmoded waveguide with resistive walls can be treated in
terms of Leontovich impedance boundary conditions and of a
bi-orthogonal set of eigenmodes, in formal analogy with those for a
microwave waveguide with perfectly conducting walls. From the
mathematical viewpoint, the iris guide theory with Vainstein
boundary conditions is not fundamentally different from the theory
of microwave waveguide with resistive walls.

\begin{figure}[tb]
\includegraphics[width=1.0\textwidth]{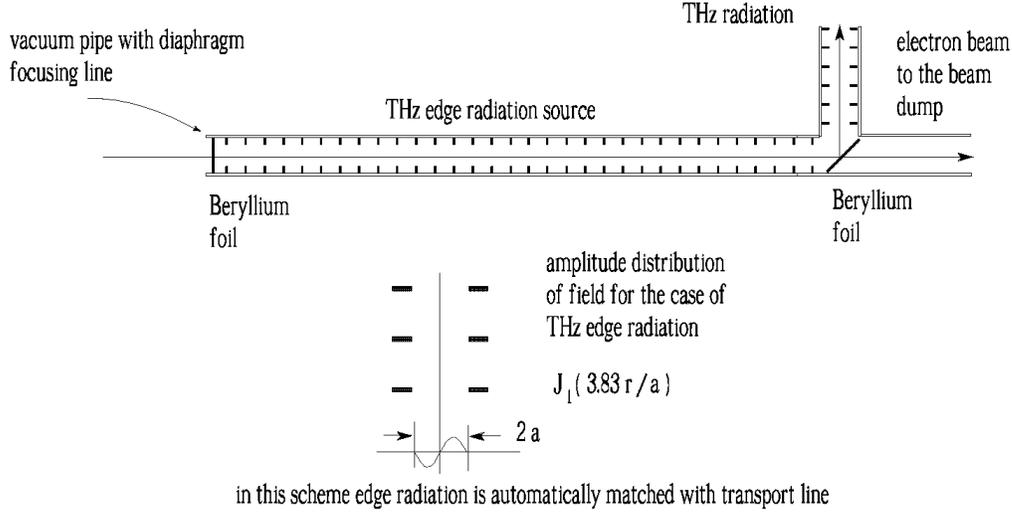}
\caption{Scheme for generating edge radiation in an iris line. In
this scheme edge radiation is automatically matched to the transport
line.} \label{th2}
\end{figure}
As already mentioned, an iris line constitutes a suitable THz beam
transport system for the LCLS baseline. Calculations indicate that
the losses of the principal mode should be in the $10\%$ level per
$100$ meters. The transfer line pipe radius is estimated to be equal
to $7$ cm. It is therefore technically feasible to install such
transfer line inside the access maze. However, in order to
efficiently couple radiation into transmission line, one also needs
to match the spatial pattern of the source radiation to the mode of
transmission line. First we analyze a coherent edge radiation source
in the THz range, which is relatively simple to implement at XFELs.
A setup where edge radiation is formed with the help of upstream and
downstream metallic screens (foils) is shown in Fig. \ref{th2}. The
edge radiation from the upstream screen is extracted by the
downstream metallic screen, which acts as a mirror, and is sent to
the iris transmission line. The length of the straight section
between upstream screen and mirror plays the role of the length of
the insertion device for edge radiation\footnote{A hole may be
present or not in the edge radiation screens. In the hard X-ray
regime of operation Be foils without hole can be used. In the soft
X-ray regime the minimal size of the hole is defined by the
condition that losses of the soft X-ray radiation from the baseline
undulator due to the aperture limitation should be avoided. In any
case, a hole with a diameter of a few mm will not perturb the soft
X-ray beam nor the electron beam, nor the THz beam.} \cite{oured}.
The matching problem is easily solved if the THz source is equipped
with an iris line as well. We consider an electron beam moving along
the z-axis inside the axisymmetric iris line, Fig. \ref{th2}. Such
source generates THz radiation with a fundamental mode that is
automatically matched to that of the transmission line. We developed
a theory of such kind of THz source using Vainstein impedance
boundary conditions. As in the case of a microwave waveguide with
resistive walls, one can use a Green's function approach to solve
the field equations. Only non-azimuthal symmetric modes turn out to
be driven by the uniform motion of the space charge distribution in
an axisymmetric iris guide. From a physical viewpoint this is sound
result. In fact, radiation is related with energy change of the
particles, which happens through the scalar product of the electric
field and velocity of the particles. Since the transverse velocity
is equal to zero in the edge radiation case, symmetric modes cannot
lead to any energy change of the electrons moving along the axis of
the axisymmetric iris  guide. Let us focus on the fundamental non
symmetric mode only. Neglecting losses, for the moment, the
amplitude for the orthogonal polarization components of the field in
a cartesian coordinate system, where $x$ and $y$ are the transverse
horizontal and vertical directions is given by:

\begin{eqnarray}
&&A_x = A_1 \cos(\phi) J_1(3.83 r/a) \cr && A_y = A_1 \sin(\phi)
J_1(3.83 r/a)~, \label{modes}
\end{eqnarray}
where $J_1$ indicates the first order Bessel function of the first
kind, $a$ is the iris radius, $A_1$ is a constant for a certain
longitudinal position, and $r = \sqrt{x^2 + y^2}$, $\phi =
\arctan(y/x)$ designate polar coordinates.  It follows from Eq.
(\ref{modes}) that the direction of the electric field is radial
i.e. varies as a function of the transverse position. Therefore, for
the edge radiation case we have two separate amplitudes for two
orthogonal polarization directions. They are not azimuthal symmetric
because they depend, respectively, on $x/r$ and $y/r$, i.e. on the
cosine and on the sine of the azimuthal angle. Only if one sums up
the intensity patterns referring to the two polarization components
one obtains the azimuthal symmetric intensity distribution

\begin{eqnarray}
I = A_1^2 J_1^2( 3.83 r/a)~. \label{inten}
\end{eqnarray}


\begin{figure}[tb]
\includegraphics[width=1.0\textwidth]{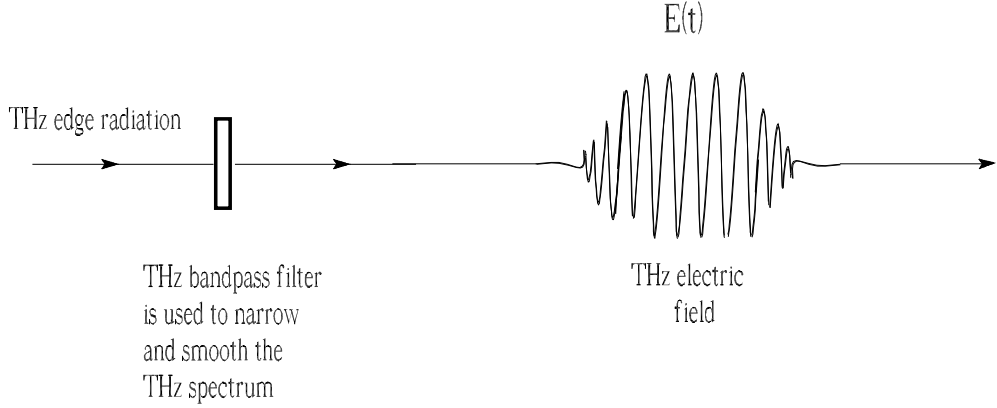}
\caption{For some experiments a smooth THz field is essential. In
these cases, a THz bandpass filter should be used.} \label{th7}
\end{figure}
Edge radiation is characterized by broad spectrum and radial
polarization. Many practical applications require the control of the
bandwidth and of the polarization of the THz pulse. The formation of
a THz edge radiation pulse usually involves monochromatization and
polarization filtering by inserting a THz band pass filter for
reducing the spectral bandwidth and a wire grid polarizer for
producing linearly polarized THz radiation, Fig. \ref{th7}.

When dealing with a setup where THz radiation from an ultra
relativistic electron beam is extracted by a mirror (in our case,
the Be foil) and sent to a THz diagnostic station or user hutch, one
usually talks about Backward Transition Radiation (BTR). The main
problem to solve is in the specification of the electric field
distribution at some position where the mirror is present. It should
be stressed that specification of the field at the mirror position
must be considered as the first step to the specification of the
field at the sample position. Such first step is considered
separately, because the field at the mirror position is independent
of the type of mirror and outcoupling optics. Once the field at the
mirror position is known, the problem of specification of the field
at the sample position can be solved with the help of Physical
Optics techniques.

Let us discuss the problem of field characterization at the mirror
position in more detail. When electrons are in unbounded space and
come from an infinitely long straight line, the field distribution
in the mirror plane can be calculated analytically following
Ginzburg and Frank \cite{GINZ}. In most practical cases, however,
the Ginzburg-Frank equation is not applicable because two basic
assumptions of the analytical derivation are not fulfilled:  the
electron beam  is moving inside the metallic vacuum chamber, which
acts effectively like an overmoded waveguide, and the straight line
has a finite length with a bending magnet at the upstream end. It is
known (see e.g. \cite{oured}) that the Ginzburg-Frank theory is the
limiting case of the more general theory of Edge Radiation (ER) in
unbounded space. Emission of edge radiation in the presence of
metallic boundaries has been a much less-treated subject in
literature, compared to the unbounded space case. To the best of our
knowledge, there is only one article reporting on edge radiation
from electrons in a homogeneous metallic  overmoded waveguide, in
particular with circular cross-section \cite{oured}. The method
described in \cite{oured} is therefore capable of treating realistic
experimental setup. Here we apply method \cite{oured} to the case of
an iris guide.

To fix ideas we focus our attention on the setup in Fig. \ref{th2}.
Electrons travel through the usual edge radiation setup, similarly
as in \cite{oured}. The difference is that now we account for the
presence of an iris guide along the straight section. Since
electrons pass through an upstream edge screen, one may assume that
the iris guide starts at the upstream screen position. In Section 5
we will calculate the field distribution at the mirror position,
which should  be subsequently propagated to the experimental hall.
The presence of an upstream edge screen seems at first glance not
necessary, because electrons come in any case from the straight
section.  However, due to variation of vacuum chamber cross-section
upstream and downstream of the baseline undulator, and presence of
the undulator magnetic field, characterization of the field
distribution at the upstream open end of the iris guide is
problematic. In spite of this, the electric field of the electron
beam  in the plane immediately behind the upstream screen can be
well defined as zero, leading to extra simplifications.  In this
case we deal with a well defined problem and this allows us to
characterize the field distribution at the end of the iris guide in
the mirror plane.

\subsection{Possible extension plans towards a coherent THz
undulator source}

\begin{figure}[tb]
\includegraphics[width=1.0\textwidth]{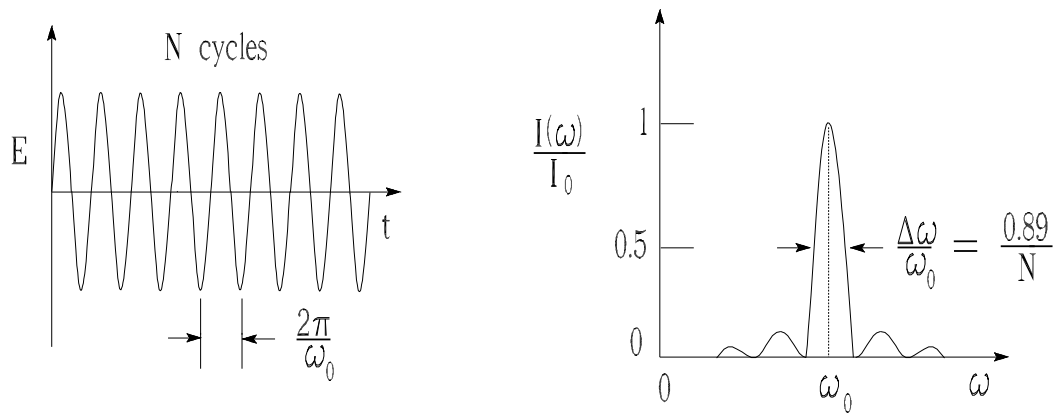}
\caption{Spectral properties of undulator radiation. A single
electron passing an undulator radiates an electromagnetic wave with
$N_w$ cycles. For on-axis radiation the relative spectral FWHM
bandwidth is $0.89/N_w$ near the central frequency.} \label{th5}
\end{figure}
\begin{figure}[tb]
\includegraphics[width=1.0\textwidth]{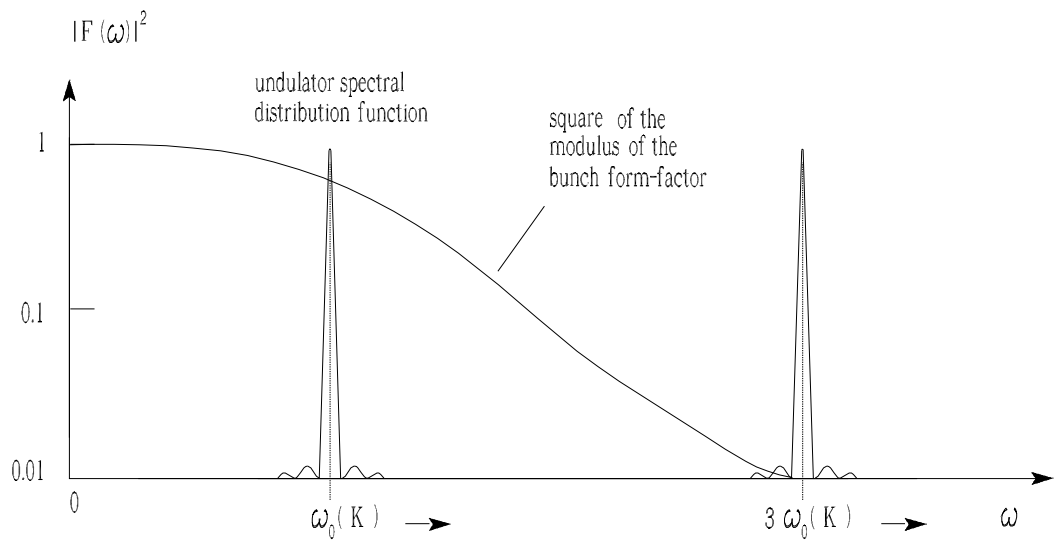}
\caption{Some typical spectra of the longitudinal electron
distribution and undulator spectral distribution function. In
general, the bunch form factor varies much more slowly in $\omega$
than the sharp undulator resonance term, and we can replace the form
factor function by the constant value at the center of the sharp
resonance curve.} \label{th6}
\end{figure}
The THz edge radiation source is limited as concerns the maximal
field strength achievable. The optimal expansion strategy for the
LCLS THz source should include a THz undulator source to be
installed at the LCLS baseline.

At resonance, undulator radiation can be described by three
parameters: period $\lambda_w$, undulator parameter $K$, and number
of periods, $N_w$. A single electron passing an undulator radiates
an electromagnetic wave with $N_w$ cycles, Fig. \ref{th5}. For the
radiation within the cone of half angle

\begin{eqnarray}
\theta_c = \frac{\sqrt{1+K^2/2}}{\gamma N_w} ~,\label{thetac}
\end{eqnarray}
the relative spectral FWHM bandwidth is $\Delta \omega/\omega =
0.89/N_w$ near the central frequency

\begin{eqnarray}
\omega_0 = 4 \pi c \frac{\gamma^2}{\lambda_w (1+K^2/2)}~,
\label{reso}
\end{eqnarray}
where $\gamma$ is relativistic factor, Fig. \ref{th5}.  The proposed
THz undulator is an inexpensive, planar electromagnetic device with
ten periods, each $1.5$ m long. At the operation wavelength of the
THz source around $0.1$ mm, the peak value of the magnetic field is
about $1$ T  at an electron beam energy around $5$ GeV.

In the case of the THz undulator at the LCLS, the electron beam
transverse size is much smaller than the diffraction size. This
means that, as pertains the characterization of the THz pulses, the
electron beam can be modeled as a filament beam. In this case the
electron beam current is made up of moving electrons randomly
arriving at the entrance of the undulator

\begin{eqnarray}
I(t) = (-e) \sum_{k=1}^{N_\mathrm{e}} \delta(t-t_k)~,
\label{current}
\end{eqnarray}
where $\delta(\cdot)$ is the Dirac delta function, $(-e)$ is the
electron charge, $N_\mathrm{e}$ is number of electrons in a bunch,
and $t_k$ is the random arrival time of the electrons at the
undulator entrance. The electron bunch profile is described by the
profile function $F(t)$. $F(t)dt$ represents the probability of
finding an electron between time $t$ and time $t+dt$. The beam
current averaged over an ensemble of bunches can then be written in
the form:

\begin{eqnarray}
\langle I(t)\rangle = (-e) N_\mathrm{e} F(t) ~. \label{Iave}
\end{eqnarray}
The radiation power at frequency $\omega$, averaged over an
ensemble, is given by the expression:

\begin{eqnarray}
\langle P(\omega) \rangle = p(\omega) [N_\mathrm{e} +
N_\mathrm{e}(N_\mathrm{e}-1)|\bar{F}(\omega)|^2]~, \label{pow}
\end{eqnarray}
where $p(\omega)$ is the radiation power from one electron and
$\bar{F}(\omega)$ is the Fourier transform of the bunch profile
function\footnote{A similar expression holds for the case of edge
radiation. In fact, the electron beam transverse size is much
smaller then the THz edge radiation diffraction size. This means
that, as pertains the characterization of the THz edge radiation
pulse, the electron beam can be modeled as a filament beam.}. For
wavelengths shorter than the bunch length the form factor reduces to
zero. For wavelengths longer than the bunch length it approaches
unity. A sample of undulator radiation spectrum is shown in Fig.
\ref{th6}. The distribution of the radiation energy within different
harmonics depends on the value of the undulator parameter $K$. In
our case of interest $K \gg 1$. Strong undulator maxima are present
at $\omega = 3 \omega_0$, $\omega = 5 \omega_0$ and so forth. The
energy measured by the detector is proportional to the convolution
of the square modulus of the bunch form factor and of the spectral
line of the undulator spectrum, as is illustrated in Fig. \ref{th6}.
Note that the bunch form factor behaves like an exponential function
at high frequencies, and falls off rapidly for wavelengths shorter
than the effective bunch length. For wavelength about three times
shorter than the effective bunch length, the radiation power is
reduced to about one percent of the maximum pulse energy at the
fundamental harmonic. As a consequence, sharp changes of the bunch
form factor result in  attenuation of the higher undulator
harmonics. In general, $|\bar{F}(\omega)|^2$ varies much slower in
frequency than the sharp resonance term, and we can replace
$|\bar{F}(\omega)|^2$ by the constant value $|\bar{F}(\omega_0)|^2$
at the center of the sharp resonance curve.

\begin{figure}[tb]
\includegraphics[width=1.0\textwidth]{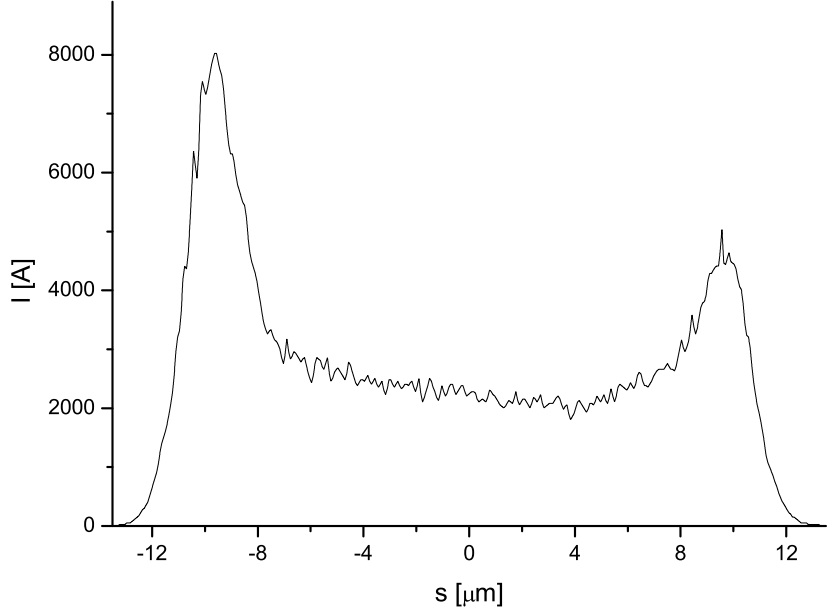}
\caption{Electron beam current profile at the LCLS, after
\cite{DING}.} \label{curr}
\end{figure}
\begin{figure}[tb]
\includegraphics[width=1.0\textwidth]{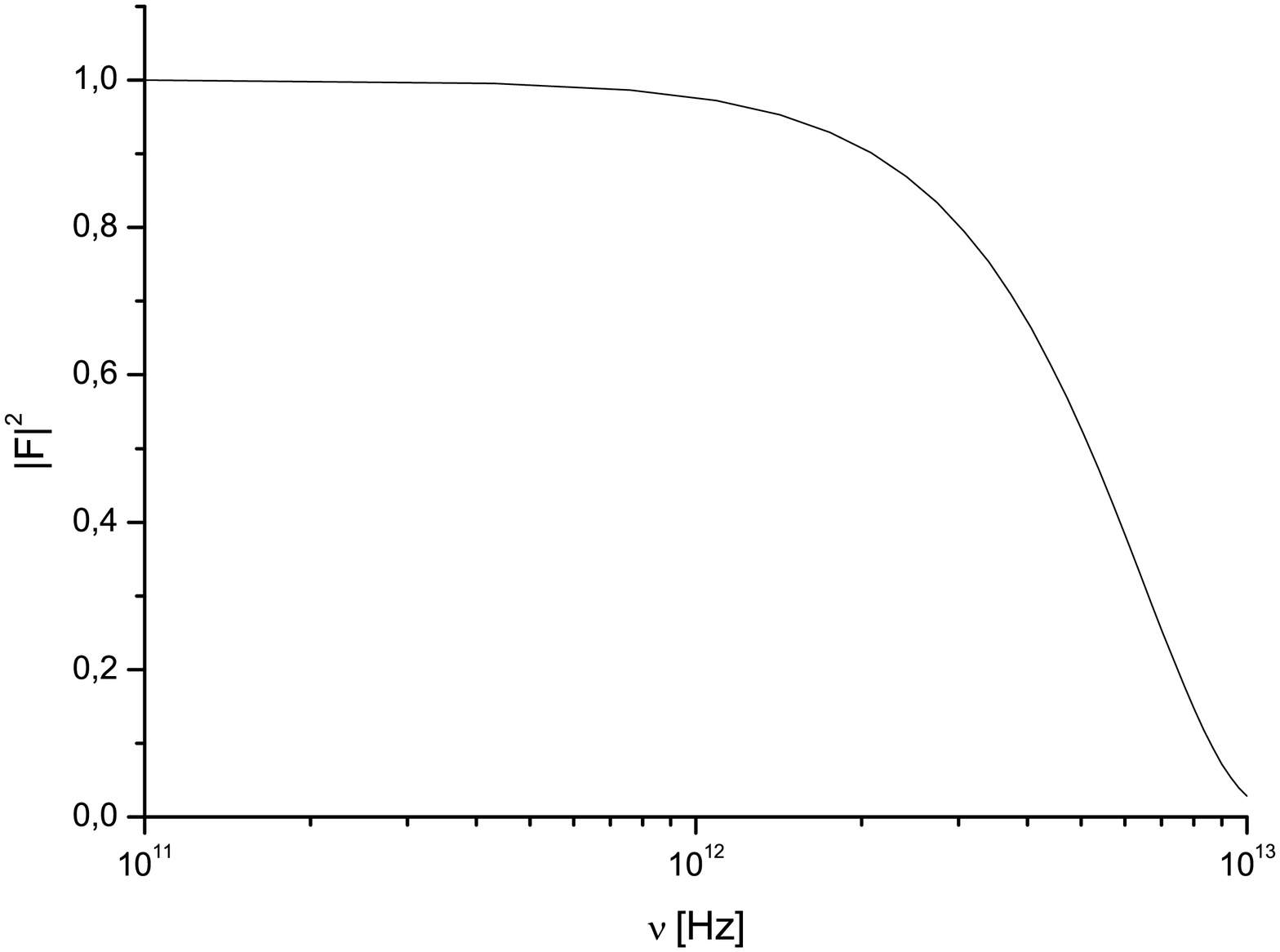}
\caption{Squared modulus of the electron beam form factor,
$|\bar{F}|^2$, corresponding to the current profile in Fig.
\ref{curr}.} \label{form}
\end{figure}
Let us consider the practically important case of an electron bunch
with strongly non-Gaussian shape similar to that used to drive the
LCLS. Fig. \ref{curr} shows the current distribution along the bunch
\cite{DING}. The nominal charge is $0.25$ nC. The electron bunch has
a complicated shape, which is reflected in the squared modulus of
the form factor shown in Fig. \ref{form}. In the case of the LCLS,
the squared of the bunch form factor modulus falls off rapidly for
wavelengths shorter than $0.06$ mm. At the opposite extreme, the
dependence of the form factor on the exact shape of the electron
bunch is rather weak and can be ignored for wavelengths longer than
$0.1$ mm.

\begin{figure}[tb]
\includegraphics[width=1.0\textwidth]{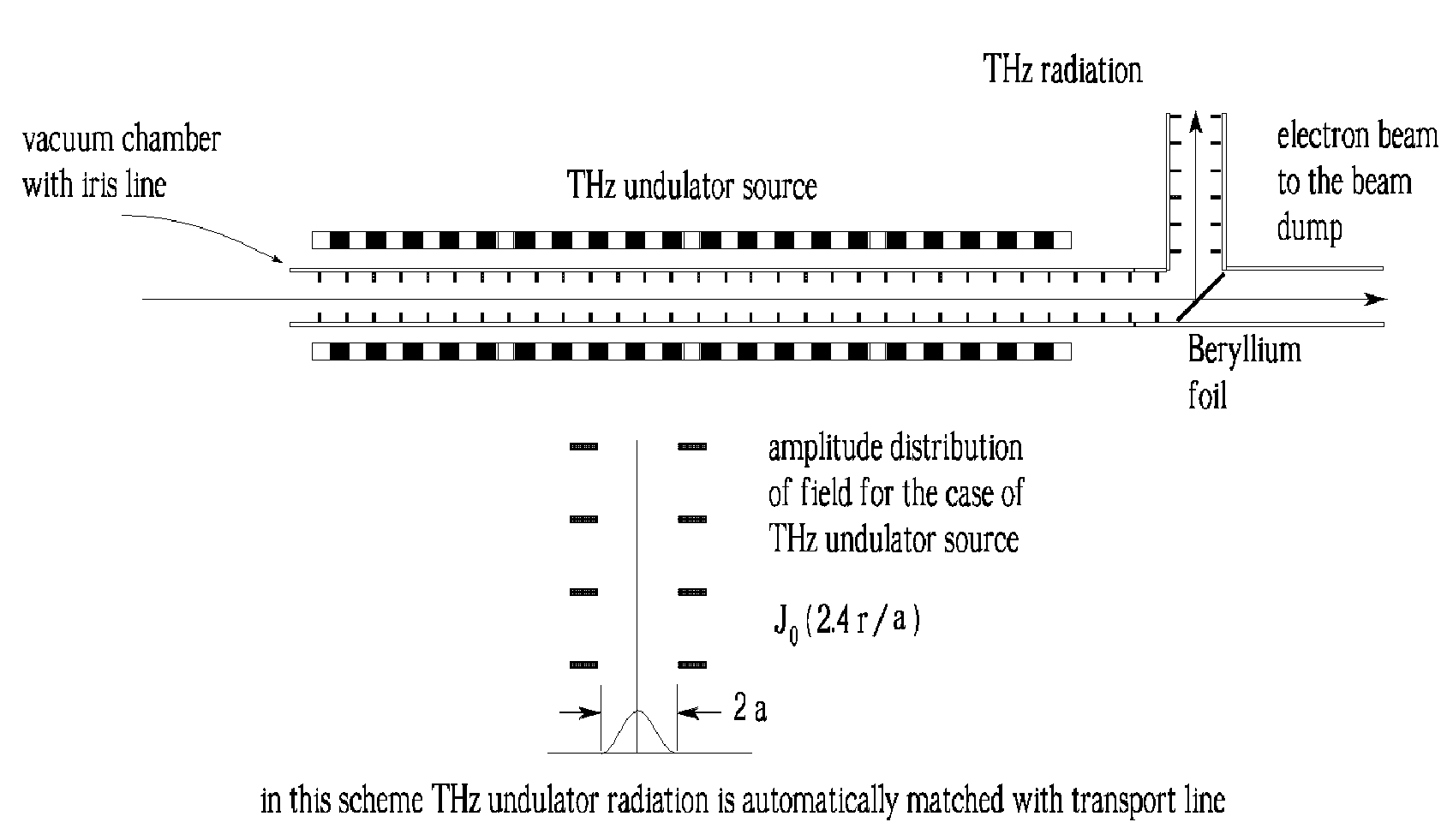}
\caption{Scheme for generating THz undulator radiation in an iris
line. The THz undulator radiation is automatically matched to the
iris transport line.} \label{th4}
\end{figure}
Similarly as for the edge radiation case, in order to solve the
matching problem with the transmission line, we propose to generate
the coherent THz undulator radiation directly in the iris line, Fig.
\ref{th4}.

Summing up, in view of the practical application to the THz
undulator line at LCLS, there is a need to develop a comprehensive
theory of undulator radiation in the presence of an iris line. As
for the edge radiation source case, one can use a Green's function
approach to solve the field equations, and the paraxial (overmoded
waveguide) approximation can be used. In addition to the paraxial
approximation, in the undulator radiation case, the resonance
approximation can be exploited too. We thus consider a large number
of undulator periods and a frequency range of interest close to the
fundamental harmonic. In free-space and under resonance
approximation the radiation from a planar undulator is horizontally
polarized, and thus constitutes a replica of the undulator
polarization properties. Moreover, the field exhibits azimuthal
symmetry. These properties are unvaried when an axisymmetric iris
line is introduced. The wiggling amplitude of the electron in the
undulator is taken to be small with respect to the dimension of the
waveguide. This greatly simplifies analytical calculations, and
describes our practical case of interest.

Applying the Green's function approach in this case one obtains a
space invariant polarization and an azimuthal symmetric field
distribution. In an axisymmetric iris guide only azimuthal symmetric
modes turn out to be driven by the wiggling electron moving along
the $z$ axis. Let us focus on the first, dominant azimuthal
symmetric mode. Neglecting losses, for the undulator radiation case
we expect the following amplitude for the horizontally polarized
radiation pulse:

\begin{eqnarray}
A_x = A_0 J_0(2.4 r/a) ~, \label{unduam}
\end{eqnarray}
where $J_0$ is the Bessel function of the first kind of zero order,
Fig. \ref{th4}.

\begin{figure}[tb]
\includegraphics[width=1.0\textwidth]{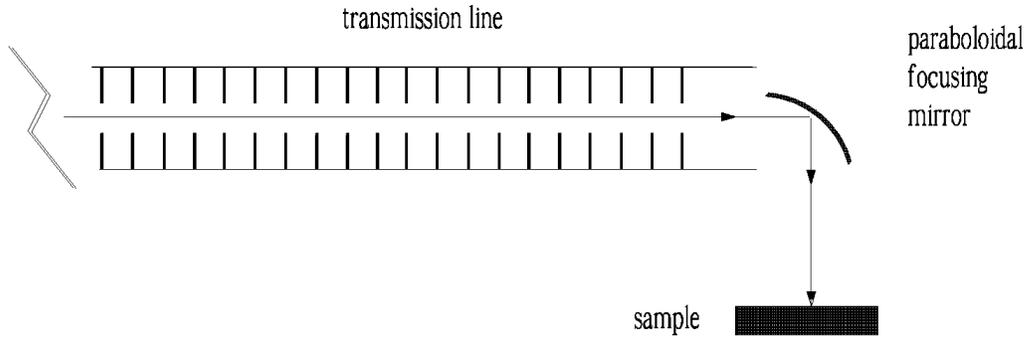}
\caption{A practical arrangement for the coherent THz source
focusing system.} \label{th14a}
\end{figure}

In contrast to edge radiation sources, THz undulator sources operate
within the typical spectral window of undulator radiation $\Delta
\omega/\omega \sim 1/N_w$ and the radiation is almost completely
linearly polarized, Fig. \ref{th5}. Additionally we should consider
the physics of the focusing process. In fact, at the end of the
transmission line the THz pulses should be focused with the help of
a converging lens or a paraboloidal focusing mirror, Fig.
\ref{th14a}. The focusing spot of the first symmetric mode of an
iris guide should be compared with the focusing spot of the first
non symmetric mode of the same iris guide. Summing up, in view of
practical applications, an important advantage of THz undulator
sources over edge radiation sources is a few ten times (for $N_w
\sim 10$)  higher intensity on the sample as compared to THz edge
radiation sources at the same electron beam parameters.

\section{\label{sec:tre} Theoretical background of an iris line}

\begin{figure}[tb]
\includegraphics[width=1.0\textwidth]{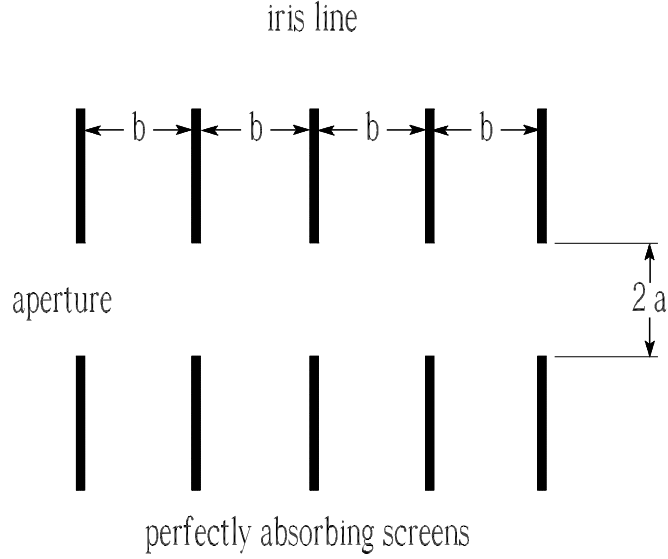}
\caption{Iris line comprising a series of collinear identical
apertures cut into parallel and equally spaced black (i.e. perfectly
absorbing) partitions of infinite extent.} \label{th9}
\end{figure}
We begin our theoretical investigation from Fig. \ref{th9}, a sketch
of an iris line, which is formed by perfectly absorbing screens of
infinite extent with collinear circular holes. The radius of the
holes is indicated with $a$, while the distance between two screens
is indicated with $b$.

An analysis of the behavior of an electromagnetic wave passing
through the line was given first in terms of numerical calculations
by Fox and Li \cite{FOX1}, and then in terms of analytical results
by Vainstein \cite{VAI1,VAI2}.

\subsection{Numerical method}

Consider the iris guide in Fig. \ref{th9}, and call
$\widetilde{E}(\vec{r'},0)$ the transverse profile of a given
polarization component of the slowly varying envelope of the
electric field on the iris at $z = 0$ in the space-frequency domain,
that is at a given fixed frequency $\omega$. The field at the iris
at $z = b$ is easily obtained as

\begin{eqnarray}
\widetilde{E}(\vec{r},b) =  \int d \vec{r'} K(\vec{r},\vec{r'};b)
\widetilde{E}(\vec{r'},0) ~, \label{Ez}
\end{eqnarray}
where function $K(\vec{r},\vec{r'};z)$ is a Green's function, or
propagator, for the paraxial wave equation which describes the field
evolution. In physical optics one uses the Fresnel propagator which
is the solution of the wave equation for unbounded space (with zero
boundary at infinity, outgoing wave). In the case of a circular
shape of the iris one may use polar coordinates and obtain
\cite{FOX1}:

\begin{eqnarray}
\widetilde{E}(r,\phi,b) =  \int_0^{2\pi} d\phi' \int_0^a dr' r'
K(r,r'\phi,\phi';b) \widetilde{E}(r',\phi',0) ~, \label{Ezpol}
\end{eqnarray}

with

\begin{eqnarray}
K(r,r'\phi,\phi';b) = \frac{i \omega \exp[-i k b]}{c b} \exp\left[-i
k \left(\frac{r^2+r'^2}{2b}-\frac{r
r'}{b}\cos(\phi-\phi')\right)\right]~, \label{Ezprop}
\end{eqnarray}
which is valid at large Fresnel numbers $a^2/(b \lambda) \gg 1$.
Iterations can be performed iris after iris, to obtain the field at
the position of any iris at arbitrary distance from the entrance.
The modes of the waveguide are, by definition, field distributions
that do not depend on the longitudinal $z$ coordinate. Since we are
dealing with the modes of an empty waveguide with no gain medium,
the mode amplitudes decreases along the $z$ axis due to diffraction
at the irises. However, the shape or spatial pattern of the field
distribution defining a given mode does not. It follows that the
field profile of a certain mode at the $n+1$ iris,
$\widetilde{E}(x,y,(n+1)b)$ is simply $\widetilde{E}(x,y,n b)$
multiplied by some complex number\footnote{Not to be confused with
the relativistic factor $\gamma$ of the electrons.} $\gamma$. It
follows that the following integral equation holds

\begin{eqnarray}
\gamma \widetilde{E}(\vec{r}) = \int d \vec{r}'
K(\vec{r},\vec{r}';b) \widetilde{E}(\vec{r}) ~, \label{inteq}
\end{eqnarray}
where for simplicity we drop the explicit reference to the $z$
dependence.

Note that Eq. (\ref{inteq}) may be written in the operator form

\begin{eqnarray}
\Gamma \widetilde{E} = \gamma \widetilde{E}~, \label{oper}
\end{eqnarray}
where $\Gamma$ is the operator corresponding to a trip between two
successive irises:

\begin{eqnarray}
\Gamma  = \int d\vec{r'} K(\vec{r},\vec{r'};b) ~ . \label{oper}
\end{eqnarray}
According  to Eq. (\ref{oper}), the modes of the iris guide are the
eigenfunctions of the operator $\Gamma$ for the guide. The number
$\gamma$ is, instead, the eigenvalue corresponding to the
eigenfunction $\widetilde{E}(\vec{r})$. In contrast with usual
quantum-mechanical situations, where eigenvalue equations include
Hermitian operator, here $\Gamma$ is generally not Hermitian. For
instance, its eigenvalues are complex, whereas the eigenvalues of a
Hermitian operator are always real.

Due to diffractive losses at the irises, the total energy associated
with the field inside the guide diminishes, and one must have
$|\gamma| < 1$. The iris guide modes may be found by solving Eq.
(\ref{Ez})  numerically. This is usually done according to the
method developed by Fox and Li in 1961. One starts by assuming some
initial field $\widetilde{E}(x,y,0)$ on an iris, usually just
$\widetilde{E}(x,y,0) = \mathrm{constant}$. The field is then
propagated to the other iris by calculating the integral in Eq.
(\ref{Ez}) numerically and iterating until the field on the irises
is unchanged (within some prescribed numerical error) on successive
iterations, except for a constant factor $\gamma$. The field yielded
by this method is a solution of Eq. (\ref{inteq}), that is, it is a
mode of iris guide. Only one mode, that is the one with smaller
losses per iteration will be returned following this procedure.
However, higher-loss modes can also be obtained. For example,
looking for solutions with the form

\begin{eqnarray}
\widetilde{E}(r,\phi,z) = R_n(r,z) \exp(- i n \phi) ~,\label{radial}
\end{eqnarray}
with $n=0,1,2...$, Eq. (\ref{Ezpol}) and Eq. (\ref{Ezprop}) can be
rewritten for $R_n(r,z)$ after integrating in $d \phi'$ as

\begin{eqnarray}
R_n(r,b) =   \int_0^a dr' r' K_r(r,r';b) R_n(r',0) ~, \label{Ezpol2}
\end{eqnarray}

with

\begin{eqnarray}
K_r(r,r';b) = \frac{i^{n+1} \omega}{c b} J_n\left(\frac{k r
r'}{b}\right) \exp\left[-i k \frac{r^2+r'^2}{2b}\right]~,
\label{Ezprop2}
\end{eqnarray}
which allows to find different modes as $n$ varies.

\subsection{Analytical method}

The analytical method by Vainstein is based on the introduction of
particular, complex boundary conditions, which are called impedance
boundary conditions. Once these boundary conditions for the field
are formulated on the virtual side surface of the iris line, the
original problem, which is about an open waveguide excitation, is
simplified to that of a closed waveguide.

\begin{figure}[tb]
\includegraphics[width=1.0\textwidth]{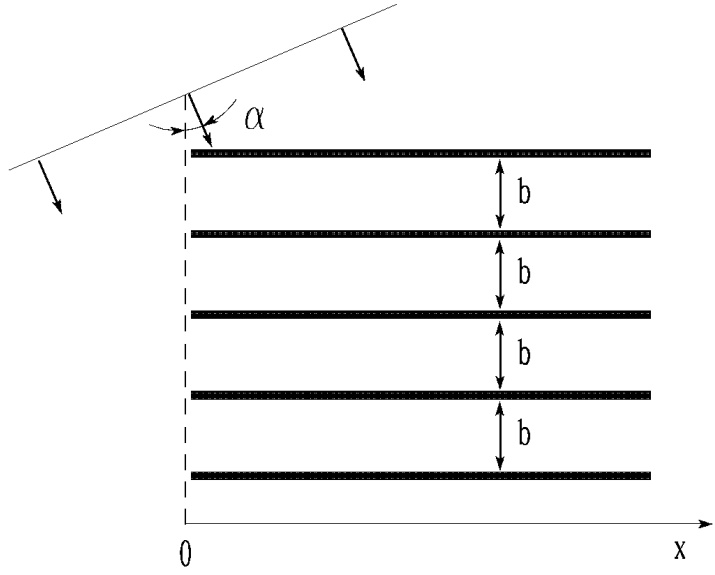}
\caption{Diffraction of plane wave at semi-infinite screens.}
\label{th10}
\end{figure}

\begin{figure}[tb]
\includegraphics[width=1.0\textwidth]{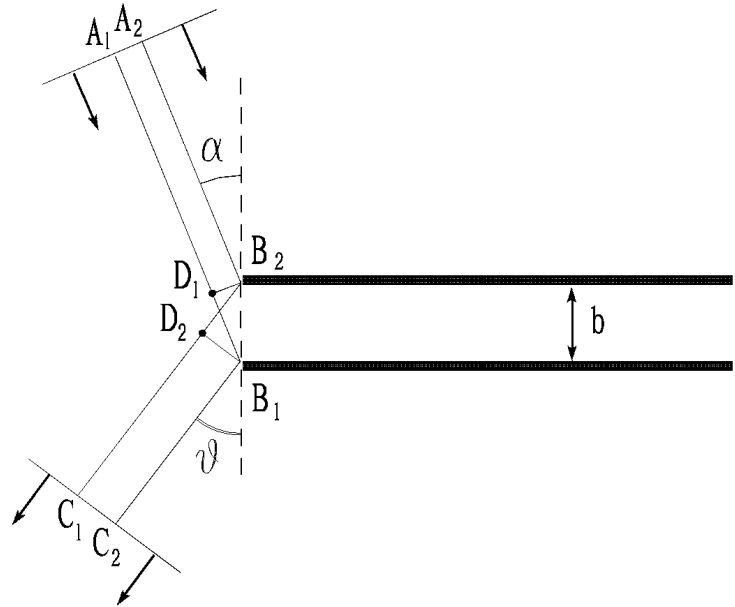}
\caption{Geometry of the diffracted wave.} \label{th11}
\end{figure}
In order to understand how the impedance boundary conditions come
into play, we consider, following \cite{VAI1,VAI2}, the simpler case
of diffraction of a plane wave at a stack of semi-infinite screens,
as depicted in Fig. \ref{th10}. We assume here that the screens are
totally absorbing. When the wave illuminates one screen, the total
field can be presented as the sum of the unperturbed wave and a
cylindrical wave generated by an image source at the edge of the
screen. The unperturbed wave propagates up to the following screen
and produces another cylindrical wave at the second edge. As a
result, the diffracted wave is given by the sum of many cylindrical
waves generated at the edges of the screens. Constructive
interference takes place only along a series of privileged
directions. This is shown in Fig. \ref{th11}. The reader can easily
recognize that the path difference between $A_1B_1C_1$ and
$A_2B_2C_2$ is just given by $D_1B_1 - D_2B_2 \simeq b
(\theta^2-\alpha^2)/2$. Constructive interference is given for

\begin{eqnarray}
k b (\theta^2-\alpha^2)/2 \simeq 2 \pi n \label{constr}
\end{eqnarray}
with $n=0, 1, 2...$, while it is zero at all other directions. For
$n=0$ one obtains the zeroth order diffracted wave, which is just
"reflected". When the angle of incidence is small $\alpha^2 \ll
\lambda/(2 \pi b)$, the screen does not absorb, but rather reflect
the wave. The finding of the reflection coefficient pertaining the
zeroth order diffraction is due to Vainstein, and reads

\begin{eqnarray}
R_0 = -\exp\left[-\beta_0 (1-i) \alpha (k b)^{1/2}\right]~,
\label{constr}
\end{eqnarray}
with\footnote{$\beta_0$ turns out to be related with one of the most
famous mathematical functions, the Riemann zeta function $\zeta$. In
fact, it is given by $-\zeta(1/2)/\sqrt{\pi}$, see
\cite{VAI1,VAI2}.} $\beta_0 = 0.824$. Note that $R_0$ is complex,
meaning that the zeroth order diffracted wave suffers phase shift
and attenuation. The impedance boundary condition are a consequence
of Eq. (\ref{constr}).

In fact, the slowly varying envelope of the total field in the plane
of the screen turns out to be written as

\begin{eqnarray}
\widetilde{E}(x) = A[\exp(i k_x x) + R_0 \exp(- i k_x x)]
~,\label{Exxxx}
\end{eqnarray}
where $A$ is just the amplitude of the incident field. Then, using
$\alpha \simeq k_x/k_z$ and $k_x^2+k_z^2 = 4\pi^2/\lambda$, and
taking advantage of Eq. (\ref{constr}) one obtains that the
logarithmic derivative of $\widetilde{E}$ at the edge of the screen
can be written as

\begin{eqnarray}
\frac{d \ln(\widetilde{E})}{dx}|_{x=0} =
\frac{1}{\widetilde{E}}\frac{d\widetilde{E}}{dx}|_{x=0} = i k_x
\frac{1-R_0}{1+ R_0}~.\label{logder0}
\end{eqnarray}
Assuming $\alpha \sqrt{kb} \ll 1$ and using the expansion
$\exp[-\beta_0 (1-i) \alpha \sqrt{kb}] \simeq 1- \beta_0(1-i)\alpha
\sqrt{kb}$, we obtain

\begin{eqnarray}
\frac{d \ln(\widetilde{E})}{dx}|_{x=0} =
\frac{1}{\widetilde{E}}\frac{d\widetilde{E}}{dx}|_{x=0} =
-\left[\beta_0 (1+i) \sqrt{\lambda
b/(8\pi)}\right]^{-1}~.\label{logder}
\end{eqnarray}
The main feature of this formula is that the term $k_x$ is
completely excluded from it, which allows one to use Eq.
(\ref{logder}) as an approximate boundary condition. Note that this
is valid not only for plane waves. In fact, any wave can be
decomposed in terms of a linear superposition of plane waves. Since
Eq. (\ref{logder}) is valid for each component, it must be valid for
their linear superposition as well. Eq. (\ref{logder}) can also be
extended for any shape of the (virtual) boundary surface, provided
that the typical value of the curvature radius is much larger than
the wavelength. Eq. (\ref{logder}) is named after Vainstein, who
first derived it, and can be written in vector form as

\begin{eqnarray}
\left[\vec{\widetilde{E}}+ (1+i) \beta_0 \sqrt{c b/(4 \omega)} ~
(\vec{n} \cdot \vec{\nabla}_\bot) \vec{\widetilde{E}}\right]_{S} =
0~, \label{vein}
\end{eqnarray}
where $S$ is the cylindrical virtual surface of the iris guide, and
$\vec{n}$ is the unit vector normal to $S$. Let us consider axially
symmetric iris line. We seek a solution for the field amplitude
$\widetilde{E}(r,\phi,z)$ in the form

\begin{eqnarray}
\widetilde{E} = u_{nj}(r)\exp[-in \phi -i k_z z]~, \label{solu}
\end{eqnarray}
with $n = 0,1, 2, ...$ . The functions $u_{nk}(r)$ are subjected to
the following homogeneous equations

\begin{eqnarray}
r^2 u_{nj}'' + r u_{nj}' + [(k_{nj})^2 - n^2)] u_{nj} = 0~,
\label{homogeq}
\end{eqnarray}

and satisfy the boundary conditions

\begin{eqnarray}
[u_{nj} + (1+i) \beta_0 \sqrt{c b/ (4\omega)} u_{nj}']_{r = a} = 0~.
\label{boundh}
\end{eqnarray}
In the first order of the small parameter $M = (8 \pi N)^{-1/2}$,
where $N = a^2/\lambda b$ is the Fresnel number with $a$ the iris
radius, the functions $u_{nj}$ assume the form

\begin{eqnarray}
u_{nj} = J_n(k_{nj} r) \label{unfun}
\end{eqnarray}
where

\begin{eqnarray}
k_{nj} = \frac{\nu_{nj}}{a} [1-(1+i) \beta_0 M]~, \label{knj}
\end{eqnarray}
and $\nu_{nj}$ is the $j$-th root of the $n$-th order Bessel
function of the first kind (i.e. $J_n(\nu_{nj}) = 0$). Substituting
the expression for $k_{nj}$ into the dispersion relation

\begin{eqnarray}
k_z^2 + k_{nj}^2 = \frac{\omega^2}{c^2} \label{kzzz}
\end{eqnarray}
we obtain the following expression for $k_z$:

\begin{eqnarray}
k_z b = \frac{\omega b}{c} - 2 \nu_{nj}^2 M^2 + 4 \nu_{nj}^2 M^3
(1+i)\beta_0~ . \label{kzzz2}
\end{eqnarray}
For the eigenmode with transverse wavenumber $k_{nj}$, the fraction
of the radiation power losses per transit of one iris is given by

\begin{eqnarray}
2 \mathrm{Im}(k_z b) = 8 \nu_{nj}^2 M^3 \beta_0 ~. \label{loss}
\end{eqnarray}
The relative loss of the $j$-th mode of order $n$ after traveling
for a distance $z$ is therefore given by

\begin{eqnarray}
\left(\frac{\Delta W}{W}\right)_{nj} = 1- \exp\left(-\frac{
\nu_{nj}^2\beta_0  }{ (2\pi N)^{3/2}}\frac{z}{b}\right) = 1-
\exp\left(-\frac{ \nu_{nj}^2\beta_0 (\lambda b)^{3/2} }{
(2\pi)^{3/2} a^3}\frac{z}{b}\right)~.\label{loss}
\end{eqnarray}
Due to the exponential dependence on $\nu_{nj}^2$, only the lower
order modes tend to survive. Note that the exponent in Eq.
(\ref{loss}) depends on the distance between two irises, $b$, only
weakly as $\sqrt{b}$, while there is a much stronger dependence on
$\lambda$ and $a$.

\subsection{Comparison between analytical and numerical methods}

\begin{figure}[tb]
\includegraphics[width=1.0\textwidth]{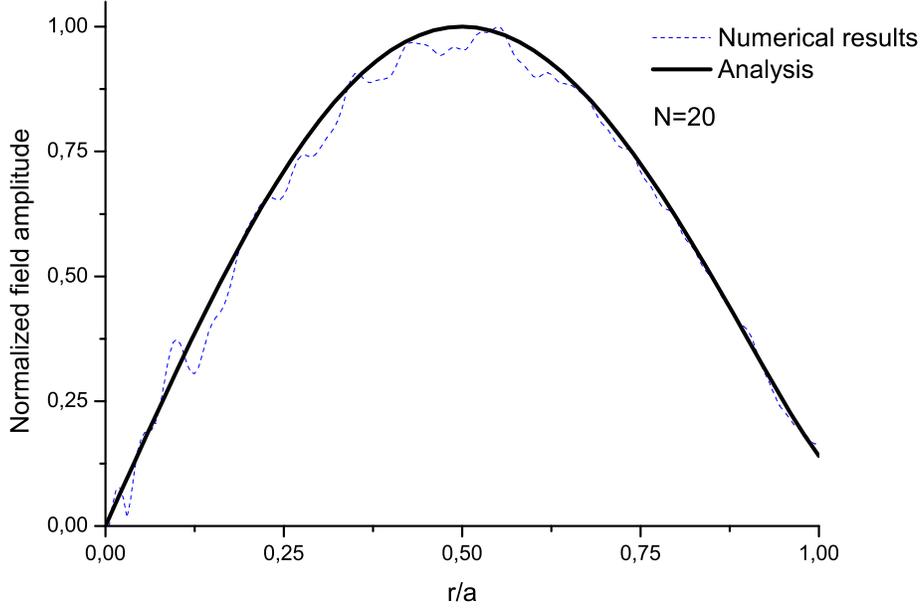}
\caption{Comparison of the fundamental non-symmetric mode derived
analytically and numerically for the Fresnel number $N =
a^2/(\lambda b) = 20$.} \label{Edshape}
\end{figure}
\begin{figure}[tb]
\includegraphics[width=1.0\textwidth]{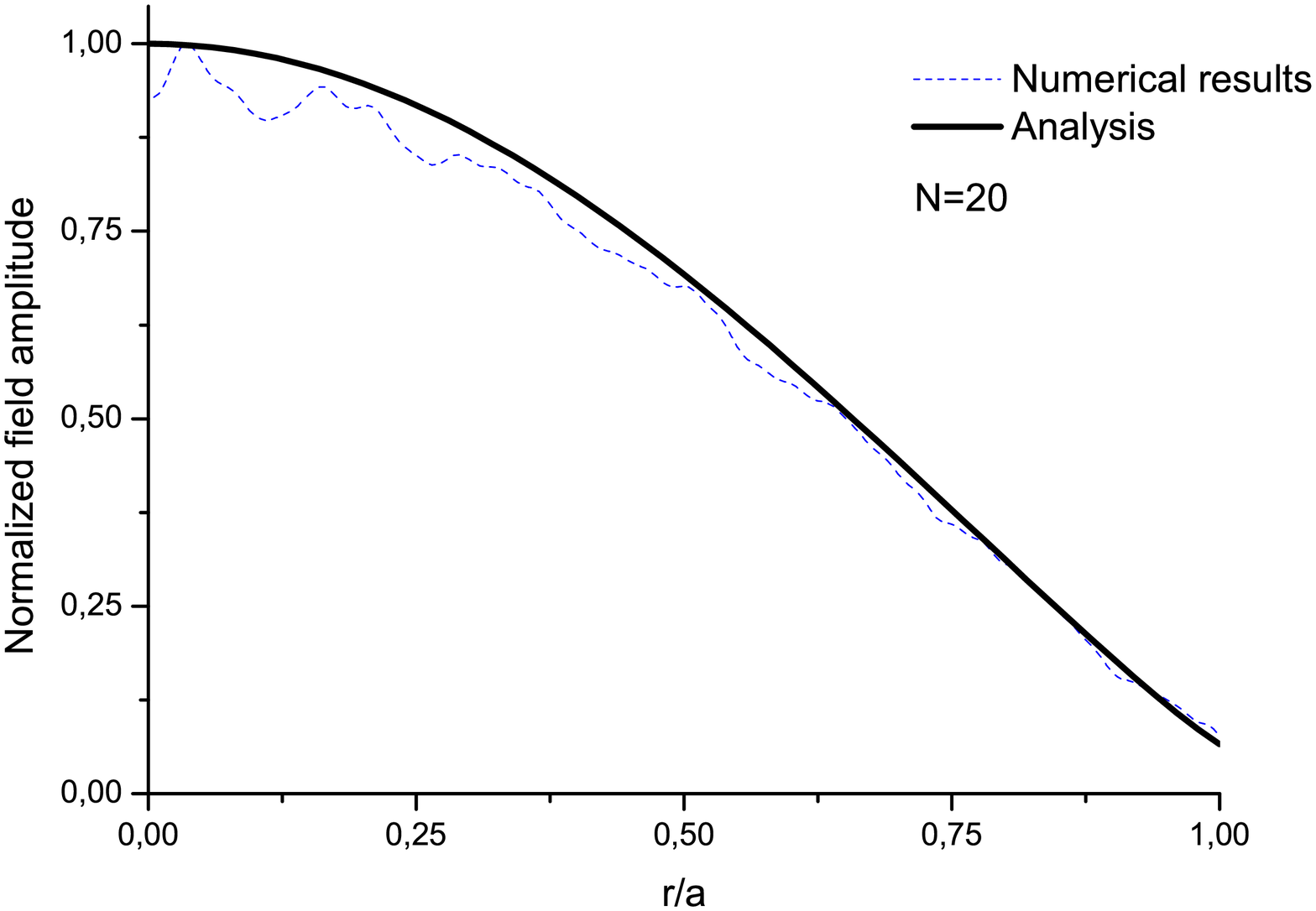}
\caption{Comparison of the fundamental symmetric mode derived
analytically and numerically for the Fresnel number $N =
a^2/(\lambda b) = 20$.} \label{Unshape}
\end{figure}
\begin{figure}[tb]
\includegraphics[width=1.0\textwidth]{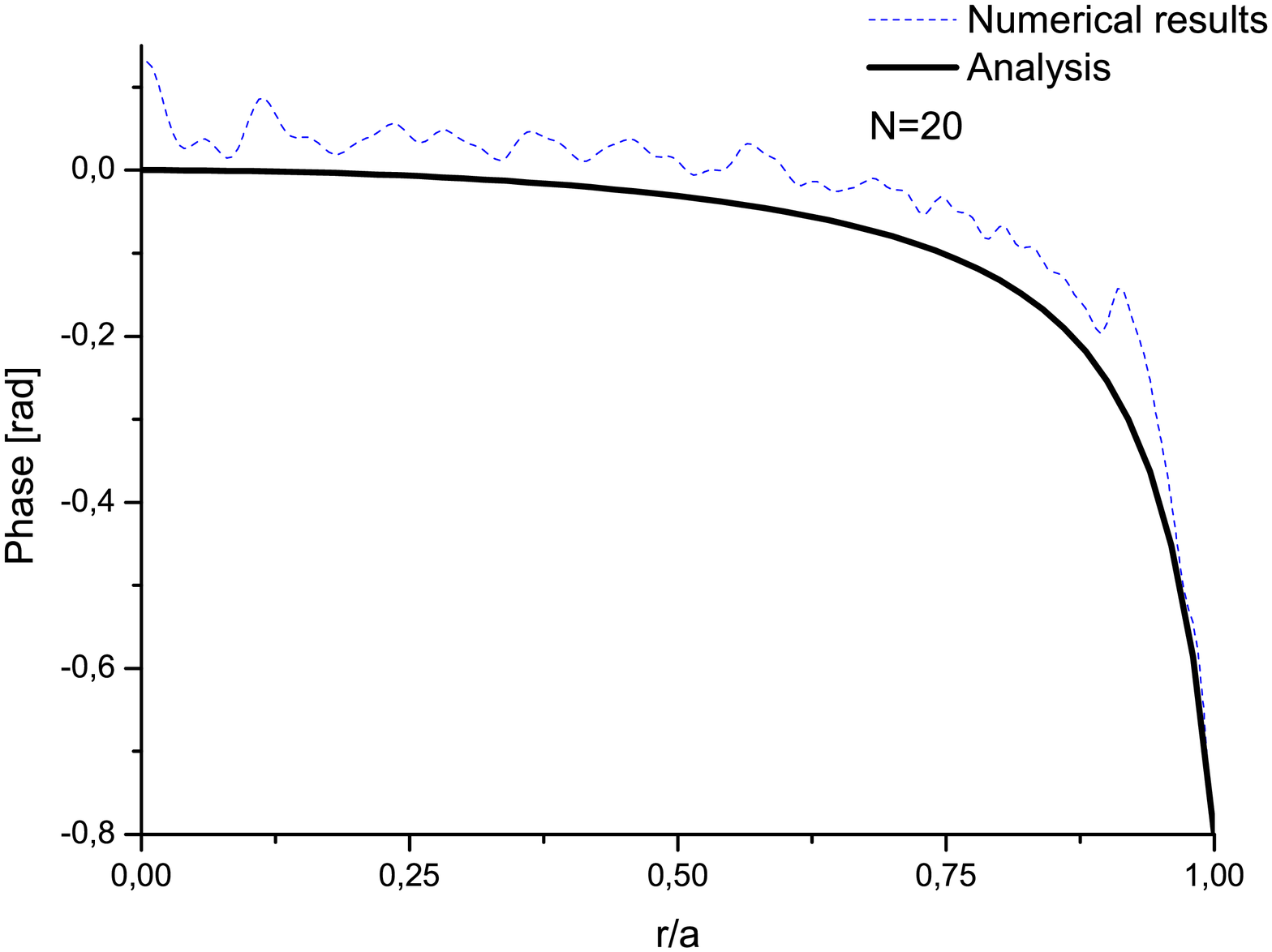}
\caption{Comparison of the phase of the fundamental symmetric mode
derived analytically and numerically for the Fresnel number $N =
a^2/(\lambda b) = 20$.} \label{Unphase}
\end{figure}
\begin{figure}[tb]
\includegraphics[width=1.0\textwidth]{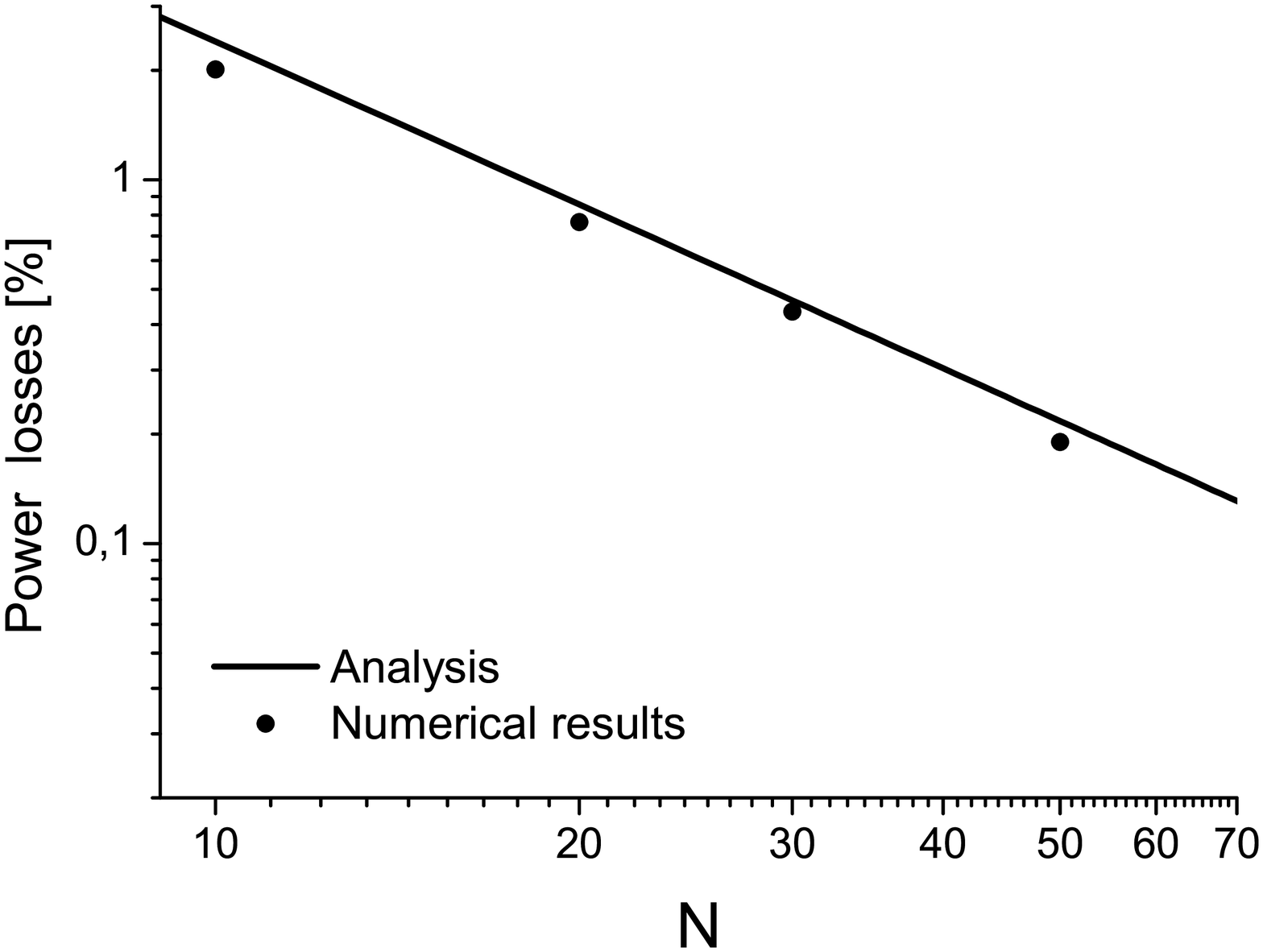}
\caption{Power losses per transit between two successive irises vs
$N = a^2/(\lambda b)$ for the first non-symmetric mode. The curve is
calculated with the analytical formula Eq. (\ref{loss}). Circles are
the result of calculations with the Fox-Li iterative method.}
\label{Edloss}
\end{figure}
\begin{figure}[tb]
\includegraphics[width=1.0\textwidth]{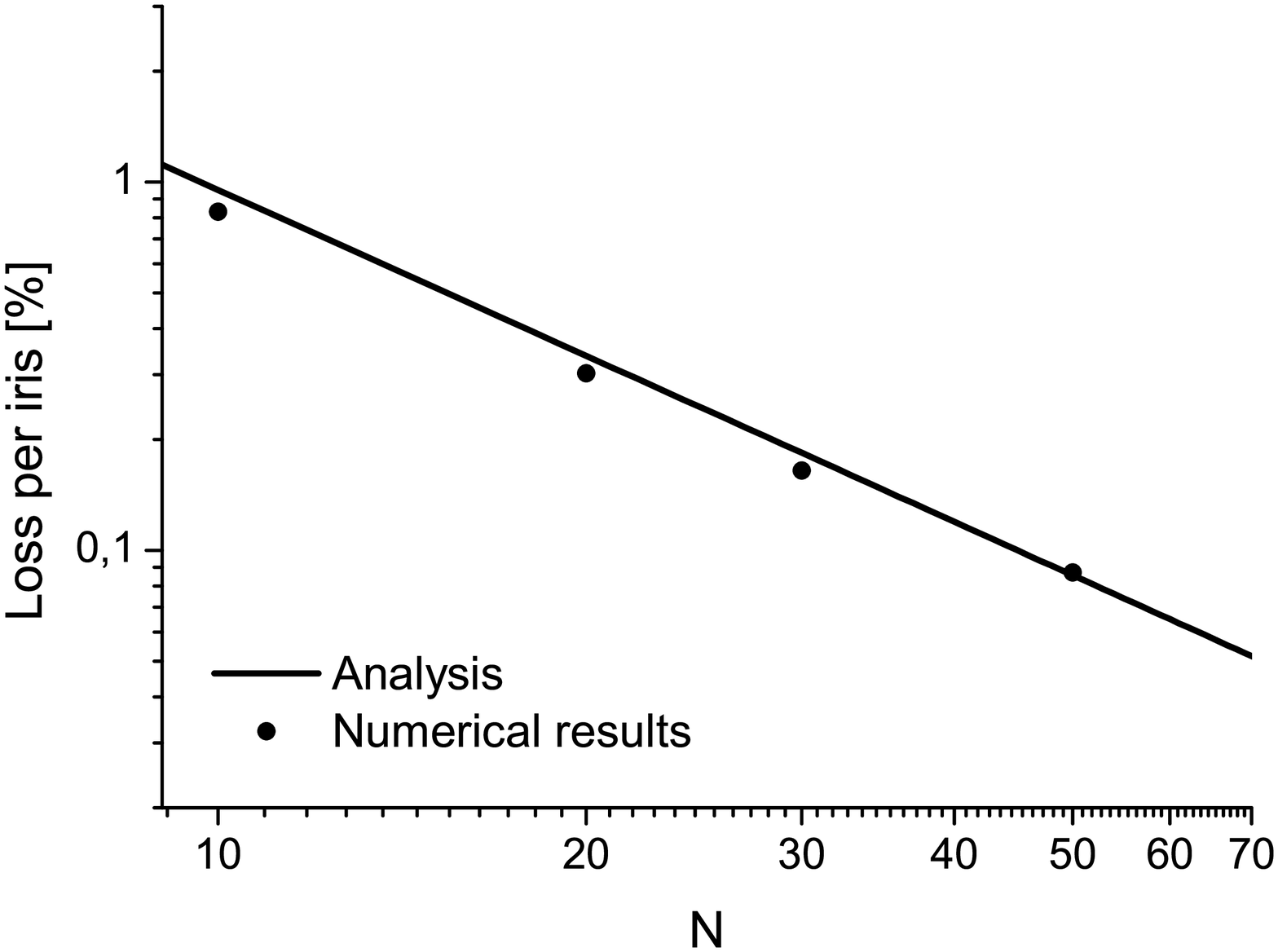}
\caption{Power losses per transit between two successive irises vs
$N = a^2/(\lambda b)$ for the first symmetric mode. The curve is
calculated with the analytical formula Eq. (\ref{loss}). Circles are
the result of calculations with the Fox-Li iterative method.}
\label{Unloss}
\end{figure}

\begin{figure}[tb]
\includegraphics[width=0.5\textwidth]{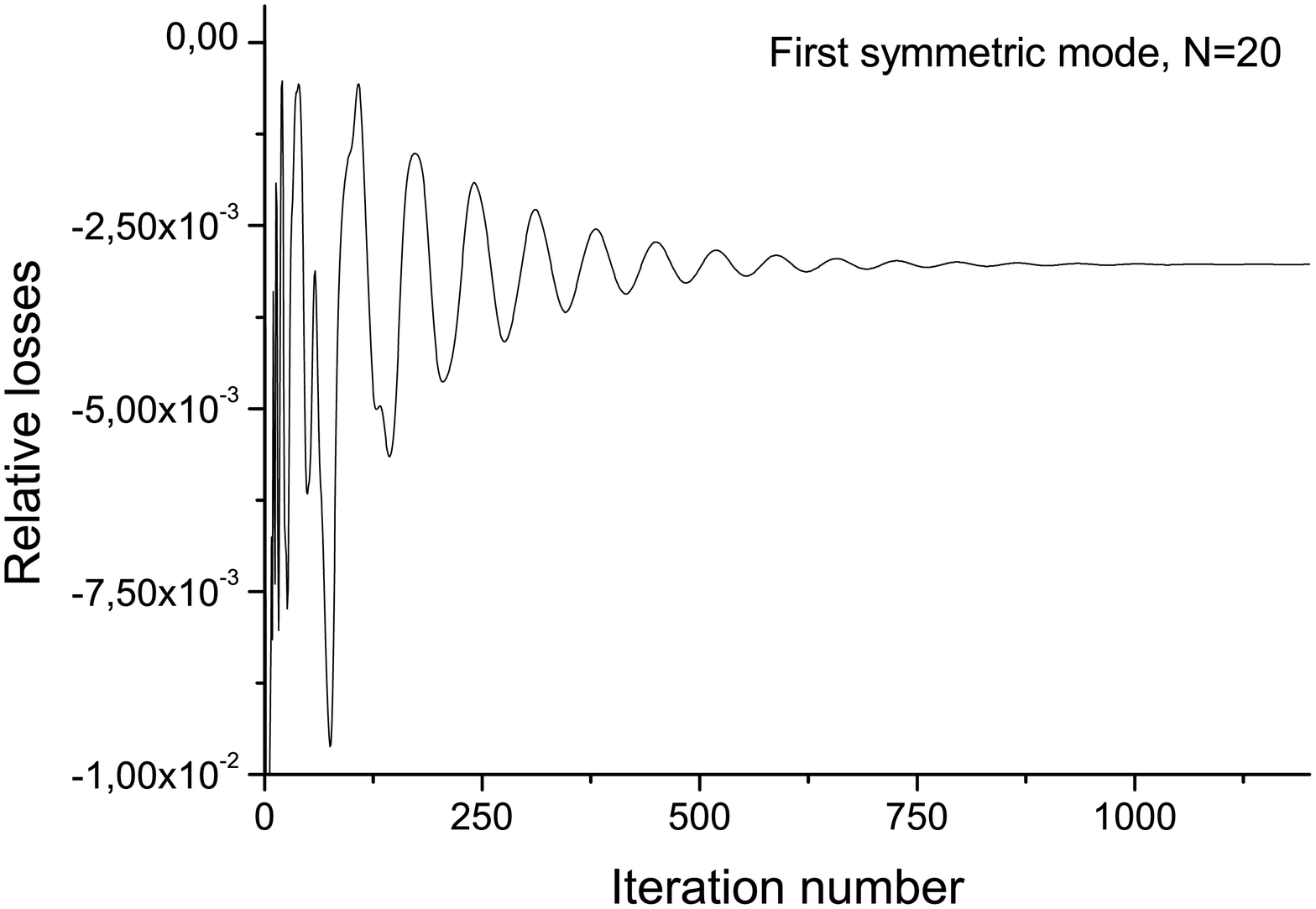}
\includegraphics[width=0.5\textwidth]{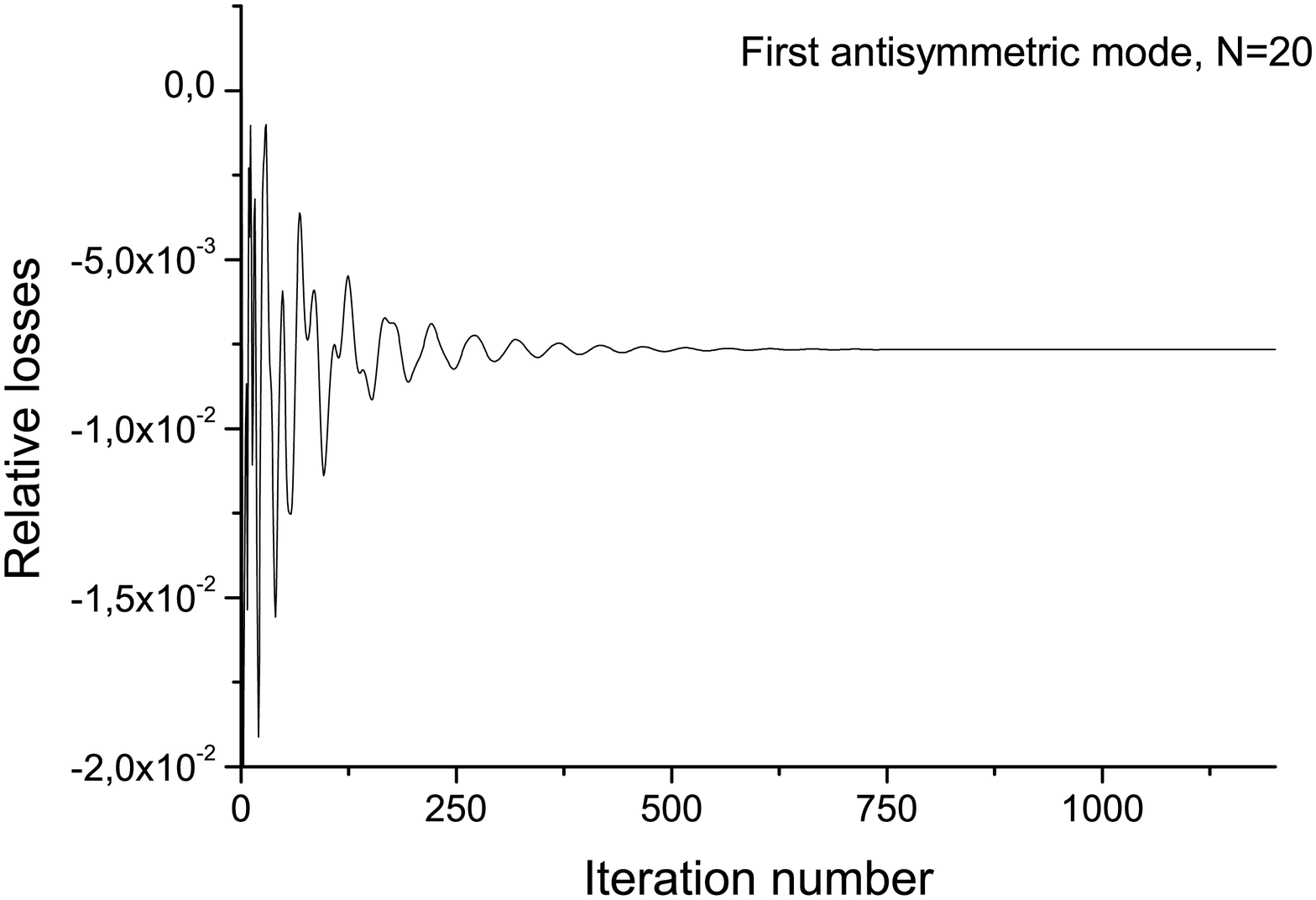}
\caption{Power losses between two successive irises at $N = 20$ as a
function of the iteration number. Left plot: first symmetric mode.
Right plot: first antisymmetric mode.} \label{conv}
\end{figure}
Comparisons between analytical and numerical results are given in
Fig. \ref{Edshape} and Fig. \ref{Unshape}, respectively for the edge
radiation and undulator cases. For the sake of completeness, Fig.
\ref{Unphase} shows a comparison between the phase of the main
symmetric mode calculated analytically and numerically as well.

Note that there are some minor differences between analytical and
numerical results. These differences, in form of ripples in the
numerically calculated modes, are not a residual from higher order
modes. In fact, they do not disappear as the number of iterations
increases. They can be explained noting that Eq. (\ref{logder}) only
accounts for the zeroth order diffraction, see Eq. (\ref{constr})
with $n=0$, while the numerical method by Fox and Li accounts for
all diffraction orders. As a result, some difference should be
expected between the iris line modes calculated with the two
methods.

A comparison between Eq. (\ref{loss}) and losses from numerical
calculations is given in Fig. \ref{Edloss} as a function of the
Fresnel number for $n=0$, and in Fig. \ref{Unloss} for $n=1$. The
convergence of these results as a function of the iteration number
is exemplified in Fig. \ref{conv} in the case $N=20$. After a large
number of iteration, the power losses between two successive irises
becomes constant. This constant is the result finally plotted in
Fig. \ref{Edloss} and Fig. \ref{Unloss}.

\subsection{Technical requirements}

\begin{figure}[tb]
\includegraphics[width=1.0\textwidth]{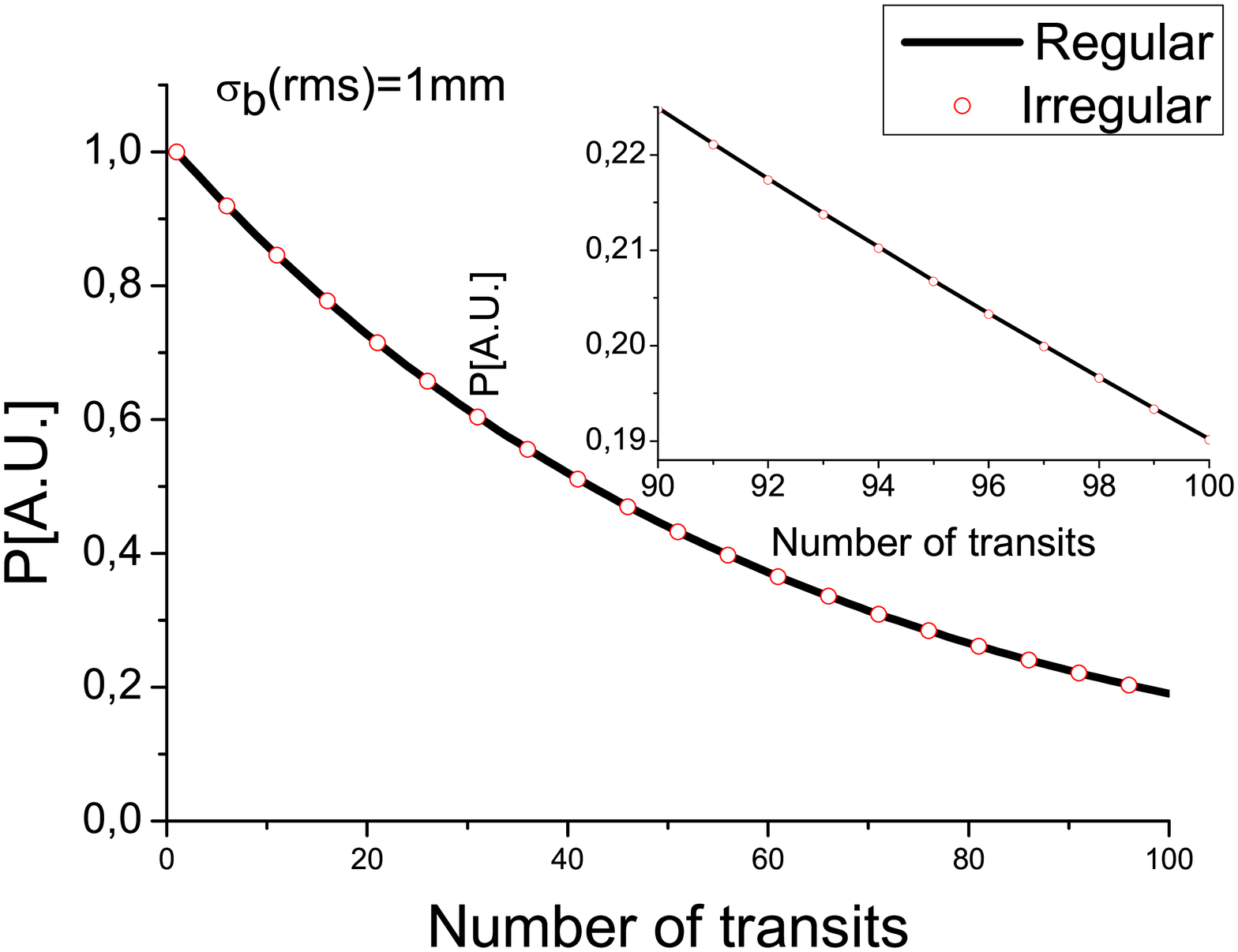}
\caption{Power of the principal symmetric mode versus number of
transits. Calculations have been performed with the Fox and Li
method. Here the wavelength $\lambda = 0.1$ mm, the iris separation
is $b = 30$ cm, the hole radius is $a = 1.35$ cm. The solid curve
corresponds to the perfect iris line. The circles represent results
from numerical simulations when distances between irises are
randomly distributed around $b = 30$ cm with rms $\sigma_b  = 1$ mm.
The inset is an enlargement of a region of the main plot. }
\label{tole}
\end{figure}

It is important to study the influence of imperfections in the
construction of the iris line on its nominal properties. With little
modification, the computer technique by Fox and Li can be used to
study the effects of misalignment of iris positions in the
longitudinal direction. The importance of this effect is shown in
Fig. \ref{tole} for $\lambda = 0.1$ mm, $b = 30$ cm, and $a=1.35$
cm. When the irises are adjusted with an accuracy of $1$ mm (rms)
the irregularities of the iris line do not result (with graphical
accuracy) in additional diffraction losses. The result that the iris
line is a "non-resonant" device should not be surprising, and can be
deduced from the explanations presented above in this Section.

Let us now study the requirements on the accuracy of the transverse
alignment between the irises.  When one of the iris is shifted off
the axis, it may cause distortion in amplitude and phase of the
diffracted wave. To estimate the change in amplitude, one should
remember that the amplitude of the wave produced by the edge of the
iris (see Fig. \ref{th10}) in the region of the next iris is
proportional to the Fresnel integral

\begin{eqnarray}
F(\psi) = \int_\psi^\infty \exp(i \tau^2) d \tau~, \label{fren}
\end{eqnarray}
where $\psi  =  x \sqrt{\pi /(\lambda b)}$.   The Fresnel integral
$F(\psi)$ is decreased by about a factor 2 when  $x$ changes from
$0$ to $\sqrt{\lambda b/\pi}$. Therefore, one can consider the
region for $x > \sqrt{\lambda b/\pi}$ as the region of the shadow.
If an iris is shifted off the axis by a value $\sim \sqrt{\lambda
b/\pi}$, either such iris or the next one fall into the region of
the shadow and does not produce any diffracted wave. To estimate
phase errors, we consider Fig. \ref{th11}. One can see that a shift
of the iris by a certain value $\delta x$ in the transverse
direction causes a phase shift of the mirror-reflected wave of about
$2 \alpha \omega \cdot \delta x/c.$ A phase shift equal to $\pi$ is
achieved at $\delta x =  \lambda /(4 \alpha)$. In paraxial
approximation $\alpha^2  \ll \lambda / (\pi b)$, and the value of
the admissible shift of the iris is mainly defined by aperture
restrictions, and not by phase distortions. Summing up, requirements
on the accuracy of the transverse alignment of the irises is given
by $\delta x \ll \sqrt{\lambda b/\pi}$. For $\lambda \simeq 0.1$ mm,
$b \simeq 30$ cm, we estimate a restriction $\delta x \ll 3$ mm.

\begin{figure}[tb]
\includegraphics[width=0.5\textwidth]{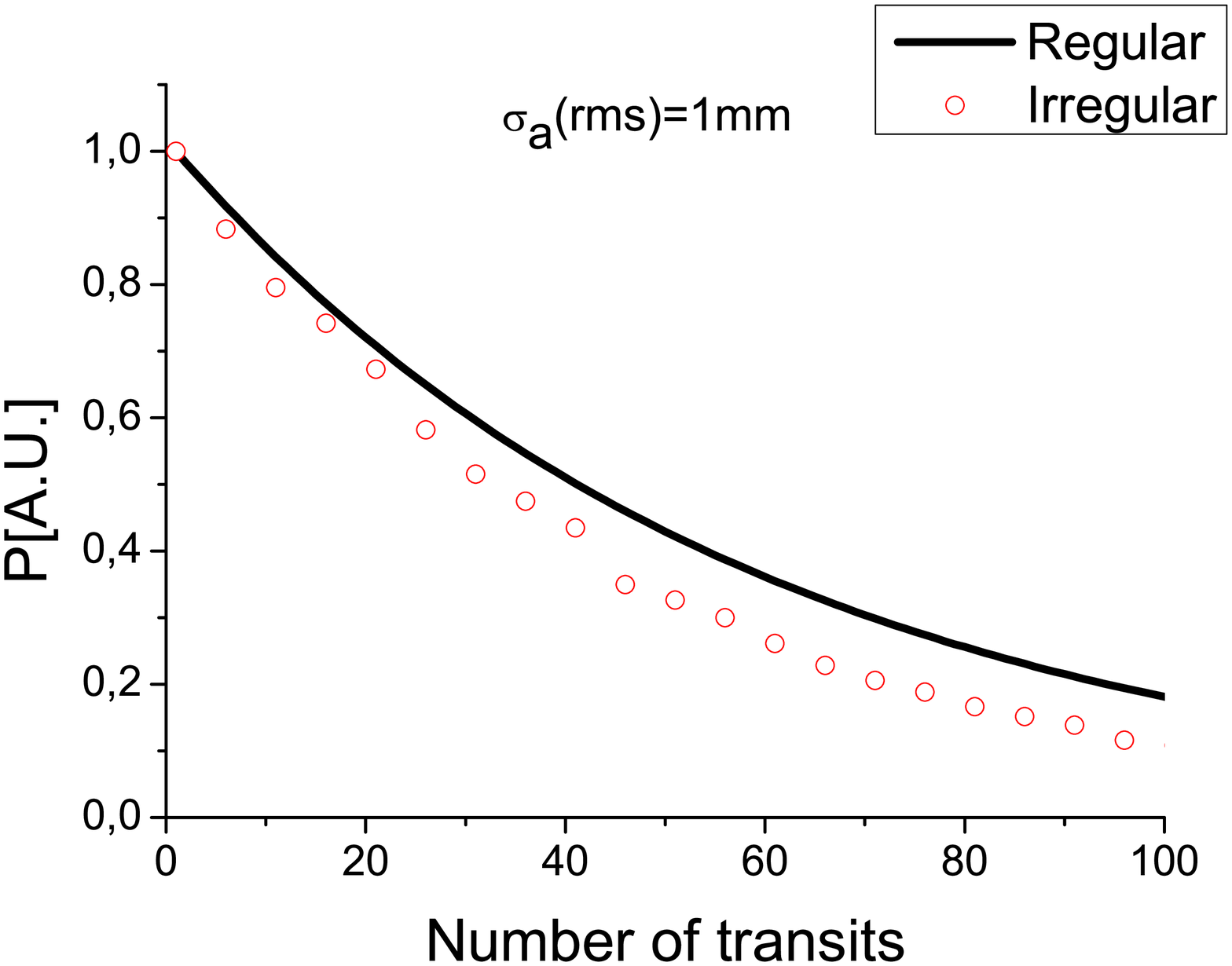}
\includegraphics[width=0.5\textwidth]{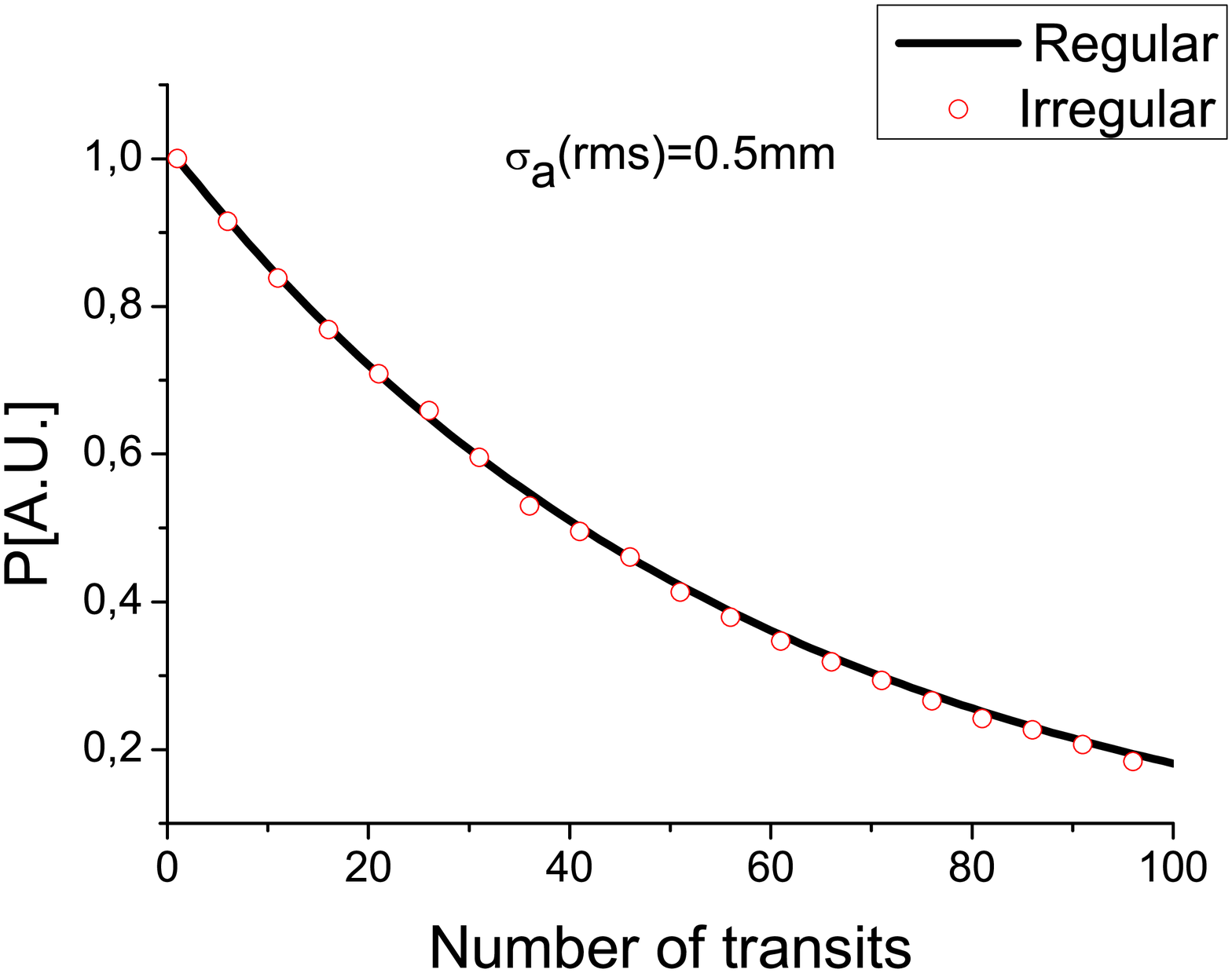}
\caption{Simulated power losses due to aperture distortions as a
function of the number of transits. The radia of irises are randomly
distributed with an rms value of $1$ mm (left plot) and $0.5$ mm
(right plot). Here the average hole radius is $a = 1.35$ cm, $b$ =30
cm, $\lambda$ = 0.1 mm. } \label{tole2}
\end{figure}

Finally, with little modifications, the computer technique employed
up to now can be used to study the effects of particular aperture
restrictions. When the radius of the irises are manufactured with an
accuracy $\delta a$, these irregularities result in extra
diffraction losses. The importance of this effect is shown in Fig.
\ref{tole2}. Simulations show that when the irises are manufactured
with an accuracy better than $1$ mm in both transverse and
longitudinal direction, manufacturing errors do not result in extra
diffraction losses in our case of interest.

In conclusion it should be noticed that requirements on the
transverse extent of the screens i.e on the difference between the
pipe radius $R$ and the hole radius $a$ is given by  $R-a  \gg
\sqrt{\lambda b/\pi} \simeq 3$ mm. Additionally, there is no need to
use completely absorbing screens in the iris line. It would even be
preferable to use reflecting screens. In fact, within the accepted
limitations, reflecting screens are almost identical to absorbing
screens as concerns diffraction effects.

\subsection{Focusing of fundamental edge radiation and undulator radiation modes}

\begin{figure}[tb]
\includegraphics[width=1.0\textwidth]{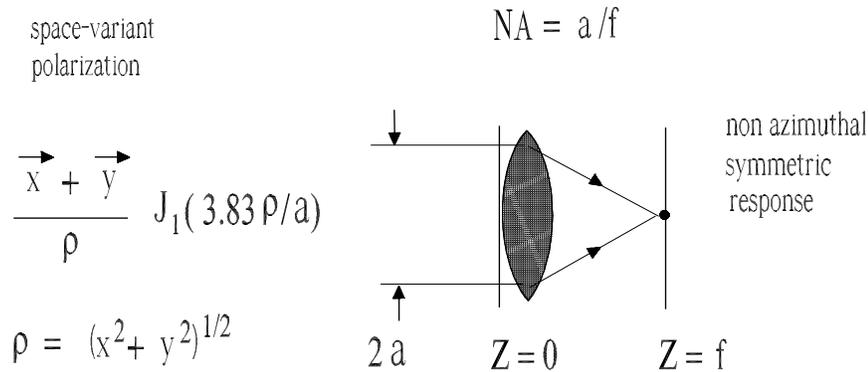}
\caption{Single lens focusing system for the edge radiation source.}
\label{th19}
\end{figure}

\begin{figure}[tb]
\includegraphics[width=1.0\textwidth]{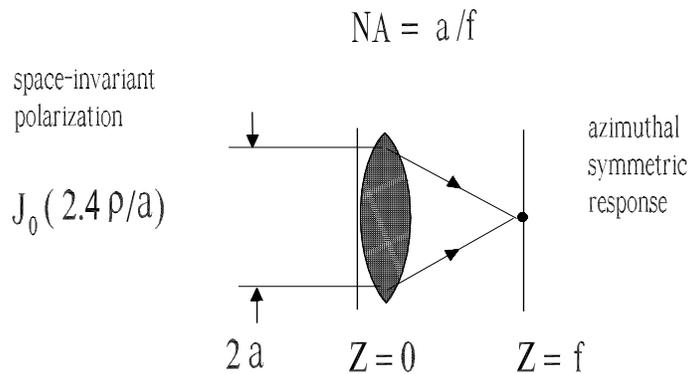}
\caption{Single lens focusing system for the undulator radiation
source.} \label{th20}
\end{figure}

For practical experimental purposes, edge or undulator THz radiation
needs to be focused onto the sample. This can be done by placing a
focusing mirror at the exit of the iris guide, and by positioning
the sample in the focus. We will treat the mirror as a thin lens
with focal length $f$. The focusing system for the edge radiation
and for the undulator radiation sources are illustrated in Fig.
\ref{th19} and Fig. \ref{th20} respectively.

In general, in the space-frequency domain, the relation between the
slowly varying envelope of the field
$\vec{\widetilde{E}}(0,\vec{r'})$ at the position immediately in
front of the lens, $z_\mathrm{lens}=0$, and the slowly varying
envelope of the field $\vec{\widetilde{E}}(f,\vec{r}_f)$ at the
sample position $z_\mathrm{sample}=f$, is given by

\begin{eqnarray}
\vec{\widetilde{E}}(f,\vec{r}_f) = \frac{i \omega}{2\pi c f}
\exp\left[\frac{i \omega r_f^2}{2 c f}\right] \int d\vec{r'}
\vec{\widetilde{E}}(0,\vec{r'}) P(\vec{r'}) \exp\left[-\frac{i\omega
\vec{r}_f \cdot \vec{r'}}{cf}\right] ~,\label{fields}
\end{eqnarray}
where $P(\vec{r'})$ indicates the pupil function, and will be
assumed here to be

\begin{eqnarray}
P(\vec{r}~) = 1  ~~\mathrm{for}~~r < a, ~P(\vec{r}~) = 0 ~~
\mathrm{otherwise}~. \label{PdefP}
\end{eqnarray}
Differences between the edge radiation and the undulator radiation
case arise due to the different expressions for
$\vec{\widetilde{E}}(0,\vec{r'})$. In particular, the radial
polarization of the edge radiation pulse leads to a non-azimuthal
symmetric response, whereas the linear polarization of the undulator
radiation pulse yields an azimuthal symmetric response.

From Eq. (\ref{modes})  we know that in the edge radiation case

\begin{eqnarray}
\vec{\widetilde{E}}(0,\vec{r'}) = A_1 J_1\left(\frac{\nu_{11}
r'}{a}\right) \left[\cos(\phi') \vec{e}_x+\sin(\phi')
\vec{e}_y\right]~,\label{ermode}
\end{eqnarray}
while from Eq. (\ref{unduam}) we know that in the undulator case the
field at the position immediately in front of the lens is

\begin{eqnarray}
\vec{\widetilde{E}}(0,\vec{r'}) = A_0 J_0\left(\frac{\nu_{01}
r'}{a}\right) \vec{e}_x~.\label{unmode}
\end{eqnarray}
Substitution of, respectively, Eq. (\ref{ermode}) and Eq.
(\ref{unmode}) in Eq. (\ref{fields}) yields the following results
for the field at the sample in the case of an edge radiation pulse

\begin{eqnarray}
\vec{\widetilde{E}}(f,\vec{r}_f) = \frac{\omega A_1}{c f a^2}
\left({\nu_{11}^2} - \frac{r_f^2 \omega^2 a^2}{c^2 f^2}\right)^{-1}
\nu_{11} J_0(\nu_{11})\exp\left[\frac{i\omega r_f^2}{2cf}\right]
J_1\left(\frac{\omega a }{c f} r_f\right)~,\label{sfielderad}
\end{eqnarray}
and of an undulator pulse

\begin{eqnarray}
\vec{\widetilde{E}}(f,\vec{r}_f) = \frac{i \omega A_0}{c f a^2}
\left({\nu_{01}^2} - \frac{r_f^2 \omega^2 a^2}{c^2 f^2}\right)^{-1}
\nu_{01} J_1(\nu_{01})\exp\left[\frac{i\omega r_f^2}{2cf}\right]
J_0\left(\frac{\omega a }{c f} r_f\right)~.\label{sfieldundu}
\end{eqnarray}
It is possible to give a  quantitative measure of how well one can
focus edge radiation pulses compared to undulator pulses. This
measure is simply given by the ratio between the maximum of
intensity in the two cases, assuming that the energy per mode per
unit spectral interval is the same.

First note that the energy per unit spectral interval per unit
surface after focusing is proportional to the squared modulus of Eq.
(\ref{sfielderad}) and Eq. (\ref{sfieldundu}), respectively for the
edge radiation and for the undulator case. The assumption of same
energy per mode per unit spectral interval can be enforced by means
of the following relation between the field amplitudes:

\begin{eqnarray}
\left(\frac{A_1}{A_0}\right)^2 =
\frac{J_1^2(\nu_{01})}{J_0^2(\nu_{11})}~, \label{A1A2}
\end{eqnarray}
where $A_0$ and $A_1$ enter in the expression for the field as in
Eq. (\ref{sfieldundu}) and Eq. (\ref{sfielderad}). The energy per
unit spectral interval per unit surface is given by

\begin{eqnarray}
\frac{d W}{d\omega dS}&& = \frac{c}{4\pi^2}
\left|\vec{\widetilde{E}}\right|^2 ~.\label{eneee}
\end{eqnarray}
Note that the transverse dependence of Eq. (\ref{sfielderad}) and
Eq. (\ref{sfieldundu}) is in the normalized distance $\hat{r}_f =
r_f \omega a / (c f)$. For the non-symmetric mode case, the maximum
of energy density in the focal plane occurs at $\hat{r}_f \simeq
2.84$. For the symmetric mode case, such maximum occurs at
$\hat{r}_f$ = 0. The relative efficiency with which the
non-symmetric mode can be focused compared with the symmetric mode
is given by $\eta \sim 1$. The outcome of this calculation is
different from what one would have intuitively guessed. If we assume
the same energy in symmetric and non-symmetric modes, after focusing
we obtain practically the same maximum energy density on the sample.
Our consideration is only valid for the aberration-free lens case
and within a paraxial treatment i.e. for numerical aperture of the
lens NA $= a/f  < 0.3$.  In fact, the field distribution of the
non-symmetric mode, Eq. (\ref{ermode}) is quite different compared
to the symmetric mode distribution Eq. (\ref{unmode}), and the
aberrations can have a significant effect on the relative mode
efficiency $\eta$.

\section{\label{sec:quattro} Green's function for an axisymmetric iris line}

Let us consider the motion of an ultrarelativistic electron in an
axisymmetric iris line, and indicate with $\vec{\bar{E}}(\vec{r}, z,
\omega)$ the Fourier transform of the transverse electric field
generated by the electron. Since the radiation formation length is
much longer than the reduced wavelength $\lambdabar = c/\omega$, the
electric field envelope $\vec{\widetilde{E}} = \vec{\bar{E}}
\exp{[-i\omega z/c]}$ turns out to be a slowly varying function of
$z$ with respect to the wavelength, and it obeys the paraxial
Maxwell equation\footnote{Here and everywhere in this paper we make
consistent use of Gaussian units.}:

\begin{eqnarray}
\mathcal{D} \vec{\widetilde{E}}(z,\vec{r}) = \vec{g}(z,\vec{r})
\label{max}
\end{eqnarray}
The differential operator $\mathcal{D}$ in Eq. (\ref{max}) is
defined by

\begin{eqnarray}
\mathcal{D} \equiv \left({\nabla_\bot}^2 + {2 i \omega \over{c}}
{\partial\over{\partial z}}\right) ~,\label{Oop}
\end{eqnarray}
where ${\nabla_\bot}^2$ is the Laplacian operator over transverse
cartesian coordinates. The vector $\vec{g}(z, \vec{r})$ is specified
by the trajectory of the source electron, $\vec{r'}(z)$, and is
written as

\begin{eqnarray}
\vec{g} = && \frac{4 \pi e}{c}   \exp\left[{i \int_{0}^{z} d \bar{z}
\frac{\omega}{2 \gamma_z^2(\bar{z}) c}}\right]
\left[\frac{i\omega}{c^2}\vec{v}(z) -\vec{\nabla}_\bot
\right]\delta\left(\vec{r}-\vec{r'}(z)\right)~.\cr && \label{fv}
\end{eqnarray}
Here we  substituted $v_z(z)$ with $c$, based on the fact that
$1/\gamma_z^2 \ll 1$. Eq. (\ref{max}) is Maxwell's equation in
paraxial approximation.

The boundary conditions are taken as in Vainstein \cite{VAI1,VAI2}
as Eq. (\ref{vein}). Since Eq. (\ref{vein}) is valid separately for
each component, we can solve Maxwell equations separately for the
field along the horizontal and the vertical directions.
Independently of the actual shape of $\vec{g}$, the full problem for
a certain polarization component can be written as

\begin{equation}
\left\{
\begin{array}{l}
\mathcal{D} {\widetilde{E}}(z,\vec{r}) = {g}(z,\vec{r})
\\
\left[{\widetilde{E}}+  (1+i) \beta_0 \sqrt{c b/(4 \omega)}~
(\vec{n} \cdot \vec{\nabla}_\bot) {\widetilde{E}}\right]_{S} = 0~.
\end{array}\right.\label{scalarp}
\end{equation}
We proceed by using a Laplace transform technique, that allows one
to dispose of the partial derivative with respect to $z$ in
$\mathcal{D}$ in the first equation in (\ref{scalarp}).

First we define Laplace $\mathcal{L}$ and inverse Laplace
$\mathcal{L}^{-1}$ transform of a function $f(z,\vec{r})$ as

\begin{equation}
\widehat{f}(p,\vec{r}) \equiv
\mathcal{L}\left[f(z,\vec{r})\right](p) = \int_0^{\infty}
f(z,\vec{r}) \exp\left[-pz\right] dz \label{ltr}
\end{equation}
with $\mathrm{Re}[p]>0$, and

\begin{equation}
f(z,\vec{r}) \equiv
\mathcal{L}^{-1}\left[\widehat{f}(p,\vec{r})\right](z) =
\int_{\delta-i \infty}^{\delta+i \infty} \widehat{f}(p,\vec{r})
\exp\left[pz\right] dp~, \label{latr}
\end{equation}
where $\delta$ is a real number larger than all the real parts of
the singularities of $\widehat{f}(p)$.

Consistently applying a Laplace transformation to the equation set
(\ref{scalarp}) we obtain a redefinition of the problem in terms of
the Laplace transform of the field ${\widehat{E}}(p,\vec{r}) \equiv
\mathcal{L}\left[{\widetilde{E}}(z,\vec{r})\right](p)$:

\begin{equation}
\left\{
\begin{array}{l}
\widehat{\mathcal{D}} {\widehat{E}}(p,\vec{r}) =
\widehat{g}(p,\vec{r}) + \frac{2 i \omega}{c}
\widetilde{E}(0,\vec{r})
\\
\left[{\widehat{E}}+  (1+i) \beta_0 \sqrt{c b/(4 \omega)} ~ (\vec{n}
\cdot \vec{\nabla}_\bot) {\widehat{E}}\right]_{S} = 0
\end{array}\right.~,\label{pbL}
\end{equation}
where $\widehat{g}$ is the Laplace transform of $g$ and

\begin{eqnarray}
\widehat{\mathcal{D}} \equiv \left({\nabla_\bot}^2 + {2 i \omega p
\over{c}}\right) ~.\label{LOop}
\end{eqnarray}
Note that the presence of the initial condition
${\widetilde{E}}(0,\vec{r})$ in Eq. (\ref{pbL}) refers to the
possibility of introducing an external field into the system. In
what follows\footnote{In our case of interest we have an edge
radiation screen at at $z = 0$, and at the position immediately
behind the screen the field of an electron is equal to zero.}
 we set ${\widetilde{E}}(0,\vec{r}) = 0$.

Now, suppose that we find a scalar function $\widehat{G}$ such that

\begin{eqnarray}
\widehat{E} = \int \widehat{G}(\vec{r}, \vec{r'},p)
\widehat{g}(\vec{r'},p) ~d\vec{r'}~. \label{Gl}
\end{eqnarray}
In this case, the inverse Laplace transform of $\widehat{G}$, that
will be simply written as ${G}$, is the Green's function for the
problem (\ref{scalarp}), inclusive of the proper boundary
conditions. As a result

\begin{eqnarray}
\widetilde{E}(\vec{r},z) = \int_{0}^{z} dz' \int d\vec{r'}~
G\left(\vec{r}, \vec{r'},z-z'\right) g\left(\vec{r'},z'\right) ~,
\label{EG}
\end{eqnarray}
where we integrate up to $z'=z$ because the radiation formation
length for $z - z'<0$ is very short with respect to the case $z - z'
>0$. Summing up,  we first have to find $\widehat{G}$, then to apply
a Laplace inverse transform in order to get $G$ and, finally, to
solve for $\widetilde{E}$.

We start by specifying the eigenvalue problem associated with the
problem set (\ref{pbL}), that is

\begin{equation}
\left\{
\begin{array}{l}
\widehat{\mathcal{D}} {F_j}(\vec{r}) = \Lambda_j {F_j}(\vec{r})
\\
\left[{F_j}+  (1+i) \beta_0 \sqrt{c b/(4 \omega)} ~ (\vec{n} \cdot
\vec{\nabla}_\bot) {F_j}\right]_{S} = 0
\end{array}\right.~.\label{pbLa}
\end{equation}
Setting by definition $\lambda_j \equiv 2i\omega p/c-\Lambda_j$
yields the following, equivalent eigenvalue problem:

\begin{equation}
\left\{
\begin{array}{l}
\nabla_\bot^2 {F_j}(\vec{r}) +  \lambda_j {F_j}(\vec{r})=0
\\
\left[{F_j}+  (1+i) \beta_0 \sqrt{c b/(4 \omega)} ~ (\vec{n} \cdot
\vec{\nabla}_\bot) {F_j}\right]_{S} = 0
\end{array}\right.~.\label{pbL2}
\end{equation}
Boundary conditions are homogeneous, so that the domain of the
Laplacian operator is the vector space of twice differentiable
(square integrable) functions obeying boundary conditions in
(\ref{pbL2}). However, the Laplacian operator defined in this way is
not self-adjoint with respect to the usual inner product definition,
$<f,g> \equiv \int_S d\vec{r} f^* g$. This is a result of the fact
that the boundary condition in Eq. (\ref{pbL2}) are not real. Then,
eigenvalues are not real, nor eigenfunctions are orthogonal with
respect to the usual inner product definition. In general we do not
know wether the spectrum is discrete, completeness is not granted
and we cannot prove the existence of a set of eigenfunctions either.
Yet, direct calculations show that

\begin{equation}
\left<{F}_{j},{{F}}_{i}\right> = \int_S {F}_j\cdot {F}_i
~d\vec{r}_\bot=\delta_{ji}~. \label{biorto}
\end{equation}
It follows that functions ${F}_j$ form a bi-orthogonal set of
eigenfunctions. This allows us to decompose $\widehat{G}$ as

\begin{equation}
\widehat{G} = \sum_j \frac{F_j F_{j}}{2 i \omega p/c-\lambda_j}~,
~\label{LGdec2}
\end{equation}
similarly as for a Sturm-Liouville problem. Note that now
eigenvalues and eigenfunctions are complex. Also note that the
dependence of $\widehat{G}$ on $p$ is at the denominator in Eq.
(\ref{LGdec2}). As a result, $\widehat{G}$ goes to zero uniformly as
$p \longrightarrow \infty$. It follows that the Mellin transform of
$\widehat{G}$, that is the Green's function $G$, can be found with
the help of Jordan's lemma and reads:

\begin{eqnarray}
G = \sum_j \left\{\mathrm{Res}\left[\widehat{G} \exp(p_j z)
\right]\right\} \label{Jor1}\end{eqnarray}
where

\begin{eqnarray}
\left(\mathrm{Res}\left[\widehat{G} \exp[p_j z] \right]\right) &=&
\lim_{p~\rightarrow -{i c \lambda_j}/({2\omega})} \left\{\left(p+
\frac{i c \lambda_j}{2\omega} \right)\widehat{G}(p) \exp[p_j
z]\right\} \cr && = \frac{c}{2i\omega} F_j\left(\vec{r}\right)
F_j\left(\vec{r'}\right)\exp\left[-\frac{i c
\lambda_j}{2\omega}z\right]\label{Jor2}
\end{eqnarray}
One therefore obtains the following expression for $G$:

\begin{eqnarray}
G = \sum_j \frac{c}{2i\omega} F_j\left(\vec{r}\right)
F_j\left(\vec{r'}\right)\exp\left[-\frac{i c
\lambda_j}{2\omega}z\right] \label{Jor3}~.\end{eqnarray}
Given the symmetry of the problem we now introduce polar coordinates
$(r, \phi)$, define $\Delta \equiv (1+i) \beta_0 \sqrt{c b/(4 \omega
a^2)}$ and rewrite Eq. (\ref{pbL2}) as

\begin{equation}
\left\{
\begin{array}{l}
\left(\frac{\partial^2}{\partial
r^2}+\frac{1}{r}\frac{\partial}{\partial
r}+\frac{1}{r^2}\frac{\partial^2}{\partial \phi^2}\right)
{F_j}(r,\phi) + \lambda_j {F_j}(r,\phi)=0
\\
\left[{F_j}+ a \Delta \partial_r F_j\right]_{r=a} = 0
\end{array}\right.~.\label{pbL3}
\end{equation}
Posing $j \equiv \{n,k\}$, we look for solutions of Eq. (\ref{pbL3})
of the form

\begin{eqnarray}
F_{j \equiv \{n,k\}} = f_{nk} \exp(-i n \phi)~, \label{BigF}
\end{eqnarray}
yielding naturally

\begin{eqnarray}
F_{j \equiv \{n,k\}} = A_{nk} J_n(\sqrt{\lambda_{nk}} r) \exp(-i n
\phi)~, \label{BigF2}
\end{eqnarray}
with

\begin{eqnarray}
J_n(\sqrt{\lambda_{nk}}a)+ a\Delta \sqrt{\lambda_{nk}}
J'_n(\sqrt{\lambda_{nk}}a)=0, \label{eigen}
\end{eqnarray}
where $J'_n$ indicates the first derivative of the Bessel function
of the first kind of order, $J_n$, with respect to the argument.
Assuming a large Fresnel number $N \equiv a^2/(\lambdabar b) \gg 1$,
which also implies $|\Delta| \ll 1$, one can solve Eq. (\ref{eigen})
for $\lambda_{nk}$ and find

\begin{eqnarray}
\lambda_{nk} = \frac{\nu_{nk}^2}{a^2}(1-\Delta)^2 \simeq
\frac{\nu_{nk}^2}{a^2}(1-2\Delta)~, \label{lambdakn}
\end{eqnarray}
where $\nu_{nk}$ is the $k$-th zero of $J_n$, that is
$J_n(\nu_{nk})=0$ for $k=1,2,3...$ .

Normalization according to the product in Eq. (\ref{biorto}),
$<F_i,F_j> = \delta_{ij}$ finally yields

\begin{eqnarray}
A_{nk}^2 \simeq \frac{1-2\Delta}{\pi a^2 J_{n+1}^2(\nu_{nk})}~.
\label{Akn}
\end{eqnarray}
The following, final expression for the Green's function $G$
follows:

\begin{eqnarray}
G = && -\frac{i c (1-2\Delta)}{2\pi \omega
a^2}\sum_{n=-\infty}^{\infty} \sum_{k=1}^{\infty} \frac{\exp[- i n
(\phi-\phi')]}{J_{n+1}^2(\nu_{nk})} \exp\left(-\frac{i c (z-z')
\nu_{nk}^2 (1-2\Delta)}{2\omega a^2}\right) \cr && \times
J_n\left(\frac{\nu_{nk}}{a}(1-\Delta) r\right)
J_n\left(\frac{\nu_{nk}}{a}(1-\Delta) r'\right)~.\label{GG}
\end{eqnarray}
It can be shown that Eq. (\ref{GG}) is in agreement with the
findings in \cite{EVG1}, where a decomposition of the field in
azimuthal components is proposed, and a Green's function is obtained
for each azimuthal component following a different reasoning. Eq.
(\ref{GG}) can now be used to find the slowly varying amplitude of
the field along any given polarization component according to Eq.
(\ref{EG}).

In order to verify the correctness of Eq. (\ref{GG}) we study the
free-space limit, that corresponds to the limit for large values of
the iris radius $a \longrightarrow \infty$.

Since we are interested in characterizing the fields over a finite
transverse direction, and since sources have a finite transverse
size, the limit $a \longrightarrow \infty$ allows one to substitute
Bessel functions in Eq. (\ref{GG}) with asymptotic expressions for
$k \gg 1$. First remember that

\begin{eqnarray}
J_n(\zeta) \approx \sqrt{\frac{2}{\pi \zeta}}
\cos\left(\zeta-\frac{\pi n}{2}-\frac{\pi}{4}\right)~,~~~~\zeta\gg
1~. \label{asympJl}
\end{eqnarray}
One sees that

\begin{eqnarray}
\nu_{nk} = \pi\left(\frac{n}{2} + k +\frac{3}{4}\right)~,
\label{munumk}
\end{eqnarray}
yielding

\begin{eqnarray}
J^2_{n+1}(\nu_{nk})\simeq \frac{2}{\pi \nu_{nk}}~. \label{JJsq}
\end{eqnarray}
Substituting Eq. (\ref{JJsq}) in Eq. (\ref{GG}), posing $\xi = \pi
k/a$ and replacing the sum over $k$ in Eq. (\ref{GG}) with an
integral over $d\xi$ we obtain:

\begin{eqnarray}
G&&\left(r,r',\phi-\phi',z-z'\right) = \frac{c}{4 i \omega \pi}
\sum_{n=-\infty}^{\infty} \int_0^\infty \d\xi \xi \exp\left[-\frac{i
c (z-z')}{2\omega  }\xi^2\right] \cr && \times J_n(\xi r) J_{n}(\xi
r')\exp[-i n(\phi-\phi')]~. \label{Gabase}
\end{eqnarray}
Eq. (\ref{Gabase}) can also be written as

\begin{eqnarray}
G&&\left(r,r',\phi-\phi',z-z'\right) = \frac{c}{4 i \omega \pi}
\int_0^\infty \d\xi \xi \exp\left[-\frac{i c (z-z')}{2\omega
}\xi^2\right] \cr && \times J_0(\xi r) J_{0}(\xi r') + \frac{c}{2 i
\omega \pi} \sum_{n=1}^{\infty} \int_0^\infty \d\xi \xi
\exp\left[-\frac{i c (z-z')}{2\omega }\xi^2\right] \cr && \times
J_n(\xi r) J_{n}(\xi r')\cos[n(\phi-\phi')]~. \label{Gabase2}
\end{eqnarray}
Here $z>z'$.  The integrals in $d \xi$ can be performed yielding

\begin{eqnarray}
G&&\left(r,r',\phi-\phi',z-z'\right) = -\frac{1}{4 \pi (z-z')}
J_0\left(\frac{\omega r r'}{c (z-z')}\right)\cr &&\times
\exp\left[\frac{i(r^2+r'^2)\omega}{2c(z-z')}\right] - \frac{1}{2
\pi} \sum_{n=1}^{\infty} i^{-n} J_n\left(\frac{\omega r r'}{c
(z-z')}\right) \cr &&\times
\exp\left[\frac{i(r^2+r'^2)\omega}{2c(z-z')}\right]
\cos[n(\phi-\phi')]~. \label{Gabase3}
\end{eqnarray}
Using the Anger-Jacobi expansion

\begin{eqnarray}
J_0(\zeta) + 2 \sum_{n=1}^{\infty} i^{-n} J_n(\zeta) \cos[n
(\phi-\phi')]=\exp[-i \zeta \cos(\phi-\phi')] \label{expesp}
\end{eqnarray}
we find that in free-space the Green's function Eq. (\ref{GG})
reduces to

\begin{eqnarray}
G = -{1\over{4\pi (z-z')}} \exp\left[ i\omega{\mid \vec{r}
-\vec{r'}\mid^2\over{2c (z-z')}}\right] \label{Gscal}~,
\end{eqnarray}
as it must be.

\section{\label{sec:cinque} Edge radiation in an iris line}

Let us consider the case of an electron emitting edge radiation in
an iris line. The vector $\vec{g}$, whose horizontal and vertical
components are needed in order to find the field according to Eq.
(\ref{EG}) can be written as

\begin{eqnarray}
\vec{g} = -\frac{4 \pi e}{c} \exp\left(\frac{i\omega z'}{2 c
\gamma^2}\right) \vec{\nabla}_\bot \delta(\vec{r})~. \label{vecgs}
\end{eqnarray}
After integrating Eq. (\ref{EG}) by parts, the horizontal and
vertical polarization components of $\vec{\widetilde{E}}$ turn out
to be given by

\begin{eqnarray}
\widetilde{E}_{x,y} = \frac{4 \pi e}{c}\int_{-L/2}^{L/2} dz'
\exp\left(\frac{i\omega z'}{2 c \gamma^2}\right)\int d\vec{r}'
\partial_{x',y'} G(\vec{r},\vec{r}',z-z') \delta(\vec{r}')~,
\label{exy} \end{eqnarray}
where the choice of the integration limits implies that the
reference system now has its origin in the center of the drift
between the two edges. Since in Eq. (\ref{GG}) the Green's function
is expressed in terms of polar coordinates, we express likewise the
partial derivatives with respect to $x'$ and $y'$ as

\begin{eqnarray}
\partial_{x'} = \cos(\phi') \partial_{r'} - \frac{\sin(\phi')}{r'}
\partial_{\phi'}
\label{parx}
\end{eqnarray}
and

\begin{eqnarray}
\partial_{y'} = \sin(\phi') \partial_{r'} - \frac{\cos(\phi')}{r'}
\partial_{\phi'}~.
\label{pary}
\end{eqnarray}
Note that one can transform the partial derivative with respect to
$y'$ to that with respect to $x'$ by just substituting $\cos(\phi')$
with $\sin(\phi')$, and $\sin(\phi')$ with $-\cos(\phi')$.
Remembering that $\delta(\vec{r'}) = \delta(r')/(\pi r')$, using

\begin{eqnarray}
\int_0^{2\pi} d\phi' \exp(-i n \phi') \cos(\phi') = \pi~
\mathrm{for}~n=\pm 1~,\mathrm{and}~ 0 ~\mathrm{otherwise}~,
\label{int1}
\end{eqnarray}
\begin{eqnarray}
\int_0^{2\pi} d\phi' \exp(-i n \phi') \sin(\phi') = \pm \pi/i~
\mathrm{for}~n=\pm 1~,\mathrm{and}~ 0 ~\mathrm{otherwise}~,
\label{int2}
\end{eqnarray}
\begin{eqnarray}
\lim_{r'\rightarrow 0} J'_{\pm
1}\left(\frac{\nu_{1k}(1-\Delta)r'}{a}\right) =\lim_{r'\rightarrow
0} \frac{J_{\pm 1}\left(\frac{\nu_{1k}(1-\Delta)r'}{a}\right)}{r'}=
\pm \frac{\nu_{1k}(1-\Delta)}{2a}~,
\end{eqnarray}
and remembering that $\nu_{nk} = \nu_{-nk}$, $J_0^2(\nu_{1k}) =
J_2^2(\nu_{-1k})$ and $J_1(x) = - J_{-1}(x)$ one finds the following
expression for the field at the mirror position:

\begin{eqnarray}
\widetilde{E}_x(r,\phi) && = -\frac{2 i e (1-\Delta) (1-2\Delta) L
\cos(\phi) }{\omega} \sum_{k=1}^{\infty}
\frac{\nu_{1k}J_1(\nu_{1k}(1-\Delta)r/a)}{a^3 J_0^2(\nu_{1k})} \cr
&& \times
{\mathrm{sinc}\left[\left(\frac{\omega}{2c\gamma^2}+\frac{c
\nu_{1k}^2(1-2\Delta)}{2\omega
a^2}\right)\frac{L}{2}\right]}\exp\left(-\frac{i c
\nu_{1k}^2(1-2\Delta)}{2\omega a^2}\frac{L}{2}\right)\label{etildex}
\end{eqnarray}
for the horizontal component and

\begin{eqnarray}
\widetilde{E}_y(r,\phi) && = -\frac{2 i e (1-\Delta) (1-2\Delta) L
\sin(\phi) }{\omega} \sum_{k=1}^{\infty}
\frac{\nu_{1k}J_1(\nu_{1k}(1-\Delta)r/a)}{a^3 J_0^2(\nu_{1k})} \cr
&& \times
{\mathrm{sinc}\left[\left(\frac{\omega}{2c\gamma^2}+\frac{c
\nu_{1k}^2(1-2\Delta)}{2\omega
a^2}\right)\frac{L}{2}\right]}\exp\left(-\frac{i c
\nu_{1k}^2(1-2\Delta)}{2\omega a^2}\frac{L}{2}\right)\label{etildey}
\end{eqnarray}
for the vertical one.

Eq. (\ref{etildex}) and Eq. (\ref{etildey}) can be used as a basis
to calculate the energy per unit spectral interval per unit surface
at the sample position. In our case of interest, the distance
between sample and extracting mirror, $L_s$, is much longer compared
with the length of the edge radiation setup. Assuming $L_s \gg L$
and accounting for coherence through the form factor
$\bar{F}(\omega)$ the field energy per unit spectral interval per
unit surface at the sample position is given by

\begin{eqnarray}
\frac{d W}{d\omega dS}&& = \frac{c}{4\pi^2} N_\mathrm{e}^2
\left|\bar{F}(\omega)\right|^2 \left|\vec{\widetilde{E}}\right|^2
\cr &&= \frac{ 4  e^2 L^2 c }{4\pi^2 a^6 \omega^2 } N_\mathrm{e}^2
\left|\bar{F}(\omega)\right|^2 \sum_{k=1}^{\infty}
\frac{\nu_{1k}^2}{J_0^4(\nu_{1k})} \exp\left(-\frac{
\nu_{1k}^2\beta_0 c^{3/2} b^{1/2} }{ \omega^{3/2} a^3}L_s\right) \cr
&& \times \mathrm{sinc}^2\left[\frac{L}{4}\left(\frac{ \omega}{c
\gamma^2}+\frac{ \nu_{1k}^2 c}{ \omega {a}^2}\right)\right]
J_1^2\left( \frac{\nu_{1k} r}{a}\right)~.\label{specend}
\end{eqnarray}
It should be noted that in the calculation of the square modulus of
the field, $\left|\vec{\widetilde{E}}\right|^2$, crossed terms with
different values of $k$ vanish. Integrating over transverse
coordinates over the iris hole, and assuming $\Delta
\omega/\omega\ll 1$ small enough to neglect the dependence on
$\omega$ in Eq. (\ref{specend}) one finds the energy of the
radiation pulse as

\begin{eqnarray}
W && =   \frac{ e^2 L^2 c }{\pi \omega a^4} N_\mathrm{e}^2
\left|\bar{F}(\omega)\right|^2 \sum_{k=1}^{\infty}
\frac{\nu_{1k}^2}{J_0^2(\nu_{1k})} \exp\left(-\frac{
\nu_{1k}^2\beta_0 c^{3/2} b^{1/2} }{ \omega^{3/2} a^3}L_s\right) \cr
&& \times \mathrm{sinc}^2\left[\frac{L}{4}\left(\frac{ \omega}{c
\gamma^2}+\frac{ \nu_{1k}^2 c}{ \omega {a}^2}\right)\right]
\frac{\Delta \omega}{\omega}~.\label{energy}
\end{eqnarray}
For the THz wavelength range and for the XFEL spent electron beam
energies, the constant $L\omega/(c \gamma^2) \ll 1$ and can be
neglected in all cases of interest.

\section{\label{sec:sei} Radiation of an electron wiggling in an iris line}

Let us now consider the case of an electron emitting radiation from
a planar undulator with period $\lambda_w$ in an iris line. We will
be interested in frequencies near the fundamental harmonic $\omega_r
= 2 k_w c \bar{\gamma}_z^2$, where $\bar{\gamma}_z =
\gamma/\sqrt{1+K^2/2}$, $k_w = 2\pi/\lambda_w$ and $K$ is the
undulator parameter. We can specify "how near" the frequency of
interest $\omega$ is compared to $\omega_r$ by introducing a
detuning parameter $C$ defined as

\begin{eqnarray}
C = \frac{\Delta \omega}{\omega_r} k_w ~,\label{Cpar}
\end{eqnarray}
where $\Delta \omega = \omega - \omega_r$. It can be shown that
within the limit

\begin{eqnarray}
|C| \ll k_w ~,\label{resonance}
\end{eqnarray}
many simplification arise and we can neglect all non-resonant
components in Maxwell equations. This amounts to neglecting the
gradient component in the vector $\vec{g}$, and to neglecting the
constrained motion $\vec{r}'(z)$ in the Dirac $\delta$-function,
that is, effectively, in the Green's function.  Following these
prescriptions, $\vec{g}$ can be written as

\begin{eqnarray}
\vec{g} = -\frac{4 \pi e}{c}  \exp\left[{i \int_{0}^{z} d \bar{z}
\frac{\omega}{2 \gamma_z^2(\bar{z}) c}}\right] \frac{i \omega}{c^2}
v_x(z) \delta(\vec{r}) \vec{e}_x~. \label{vecgsu}
\end{eqnarray}
Eq. (\ref{EG}) yields the following expression for the horizontal
field component $\widetilde{E}$:

\begin{eqnarray}
\widetilde{E} = \frac{e \omega \theta_s
A_{JJ}}{c^2}\int_{-L_w/2}^{L_w/2} dz' \exp\left(i C z'\right)
\int_{0}^{a} d r' \int_0^{2\pi} d\phi' G(r,r',\phi,\phi',z-z')
\delta(r') ~,\label{Eu} \end{eqnarray}
where $\theta_s = K/\gamma$, $L_w$ is the undulator length and
$A_{JJ} = J_0[K^2/(4+2K^2)]-J_1[K^2/(4+2K^2)]$. Inspection of Eq.
(\ref{GG}) shows that only the term with $n=0$ yields non-zero
contributions. One therefore obtains the following
azimuthal-symmetric expression for the field at the exit of the
undulator:

\begin{eqnarray}
\widetilde{E}(r) &&=- \frac{ i  (1-2\Delta)  e  \theta_s A_{JJ}}{c
a^2}\sum_{k=1}^{\infty} \frac{1}{ {J_{1}^2(\nu_{0k})}}
J_0\left(\frac{\nu_{0k}}{R}(1-\Delta) r\right) \exp\left(-\frac{i c
\nu_{0k}^2 (1-2\Delta)}{2\omega a^2} \frac{L_w}{2}\right)\cr &&
\times \int_{-L_w/2}^{L_w/2} dz' \exp\left(i C z'\right)
\exp\left(\frac{i c \nu_{0k}^2 (1-2\Delta) z'}{2\omega a^2}\right)
~.\label{Eu2}
\end{eqnarray}
Performing the integration in $dz'$ finally yields

\begin{eqnarray}
\widetilde{E}(r) &&=- \frac{ i  (1-2\Delta)  e  \theta_s A_{JJ}}{c
a^2}\sum_{k=1}^{\infty} \frac{1}{ {J_{1}^2(\nu_{0k})}}
J_0\left(\frac{\nu_{0k}}{R}(1-\Delta) r\right) \exp\left(-\frac{i c
\nu_{0k}^2 (1-2\Delta)}{2\omega a^2}\frac{L_w}{2}\right)\cr &&
\times L_w \mathrm{sinc}\left[\frac{L_w}{2}\left(C+\frac{c
\nu_{0k}^2 (1-2\Delta) }{2\omega a^2}\right)\right] ~.\label{Eu3}
\end{eqnarray}
Eq. (\ref{Eu3}) can be used as a basis to calculate the energy per
unit spectral interval per unit surface at the sample position. In
our case of interest, the distance between sample and extracting
mirror, $L_s$, is much longer compared with the length of the
undulator. Assuming $L_s \gg L_w$ and accounting for coherence
through the form factor $\bar{F}(\omega)$, the field energy per unit
spectral interval per unit surface at the sample position is given
by

\begin{eqnarray}
\frac{d W}{d\omega dS}&& = \frac{c}{4\pi^2} N_\mathrm{e}^2
\left|\bar{F}(\omega)\right|^2 \left|\vec{\widetilde{E}}\right|^2
\cr &&= \frac{e^2 L_w^2 \theta_s^2 A_{JJ}^2}{ 4 \pi^2 c {a}^4}
N_\mathrm{e}^2 \left|\bar{F}(\omega)\right|^2 \sum_{k=1}^{\infty}
\frac{1}{J_{1}^4(\nu_{0k})}\exp\left(-\frac{ \nu_{0k}^2\beta_0
c^{3/2} b^{1/2} }{ \omega^{3/2} a^3} L_s \right)\cr && \times
\mathrm{sinc}^2\left[ \frac{L_w}{2}\left(C +\frac{c \nu_{0k}^2}{2
a^2 \omega } \right)\right] J_0^2\left(\nu_{0k} \frac{r}{a}\right)
~,\label{specend2}
\end{eqnarray}
It should be noted that in the calculation of the square modulus of
the field, $\left|\widetilde{E}\right|^2$, crossed terms with
different values of $k$ vanish. Integrating over transverse
coordinates over the iris hole, and assuming $\Delta \omega/\omega
\sim 1/N_w \ll 1$ small enough to neglect the dependence on $\omega$
in Eq. (\ref{specend2}) one finds the energy per spectral interval
$\Delta \omega/\omega$ as

\begin{eqnarray}
W&& = \frac{e^2 L_w^2 \omega \theta_s^2 A_{JJ}^2}{ 4 \pi c {a}^2}
N_\mathrm{e}^2 \left|\bar{F}(\omega)\right|^2 \sum_{k=1}^{\infty}
\frac{1}{J_{1}^2(\nu_{0k})}\exp\left(-\frac{ \nu_{0k}^2\beta_0
c^{3/2} b^{1/2} }{ \omega^{3/2} a^3} L_s\right)\cr && \times
\mathrm{sinc}^2\left[ \frac{L_w}{2}\left(C +\frac{c \nu_{0k}^2}{2
a^2 \omega } \right)\right]  \frac{\Delta
\omega}{\omega}~.\label{energyw}
\end{eqnarray}
As expected, different modes do not interfere. The total pulse
energy is given by the sum of the energy of each of the empty iris
guide eigenmode excited within the open waveguide.

One figure of merit of interest is the ratio between the pulse
energy generated  and transported  within the iris guide, and the
pulse energy  generated and transported in free space. In the
space-frequency domain, the field distribution from the undulator in
free space is easily derived. One should take the limit for $a
\longrightarrow \infty$ in Eq. (\ref{Eu}).  Accounting for Eq.
(\ref{Gscal}), we get the following well-known expression in the far
zone limit (i.e. for $z \gg L_w$):

\begin{eqnarray}
{\vec{\widetilde{E}}}_\bot&&(z, \vec{\theta}) = -\frac{\theta_s
\omega e L_w A_{JJ} }{2 c^2 z} \exp\left[i\frac{\omega \theta^2
z}{2c}\right] \mathrm{sinc}\left[\frac{L_w}{2}\left(C+\frac{\omega
\theta^2}{2c} \right)\right] \vec{e}_x~,\cr && \label{generalfin4}
\end{eqnarray}
where $\vec{\theta} = \vec{r}/z$ defines the observation direction,
and $\theta = |\vec{\theta}|$. The distribution of energy radiated
per unit solid angle, per unit frequency interval is defined as

\begin{eqnarray}
\frac{d W}{d \omega d \Omega} = \frac{c z^2}{4\pi^2} N_\mathrm{e}^2
\left|\bar{F}(\omega)\right|^2 \left|\vec{\widetilde{E}}\right|^2 ~.
\label{distrWoO}
\end{eqnarray}
Integration over angles can be easily performed  leading to

\begin{eqnarray}
W_{\mathrm{fs}} = \frac{e^2 L_w \omega^2 \theta_s^2 A_{JJ}^2}{8 c^2}
N_\mathrm{e}^2 \left|\bar{F}(\omega)\right|^2 \frac{\Delta
\omega}{\omega}~.\label{fsW}
\end{eqnarray}
This expression corresponds to the particular choice of detuning
parameter $C = 0$, at which the pulse energy  (in free space)
achieves its maximum. The ratio between the energy of the pulse
generated and transmitted with quasi-optical techniques and that of
the pulse generated and transmitted in free space provides a measure
of the departure from ideal performance. Accounting for Eq.
(\ref{loss}), Eq. (\ref{energyw}), and Eq (\ref{fsW}) we get the
following expression for this ratio:

\begin{eqnarray}
\frac{W}{W_{\mathrm{fs}}} &&= \frac{2}{\pi} \frac{c L_w}{\omega a^2}
\sum_{k=1}^{\infty} \frac{1}{J_{1}^2(\nu_{0k})}\exp\left[-\frac{
\nu_{0k}^2\beta_0 c^{3/2} b^{1/2} }{ \omega^{3/2} a^3} L_s\right]\cr
&& \times \mathrm{sinc}^2\left[ \frac{L_w}{2}\left(C +\frac{c
\nu_{0k}^2}{2 a^2 \omega } \right)\right]~. \label{finee}
\end{eqnarray}

\section{\label{sec:sette} Scheme for generating THz edge radiation from the LCLS baseline}

\begin{figure}[tb]
\includegraphics[width=1.0\textwidth]{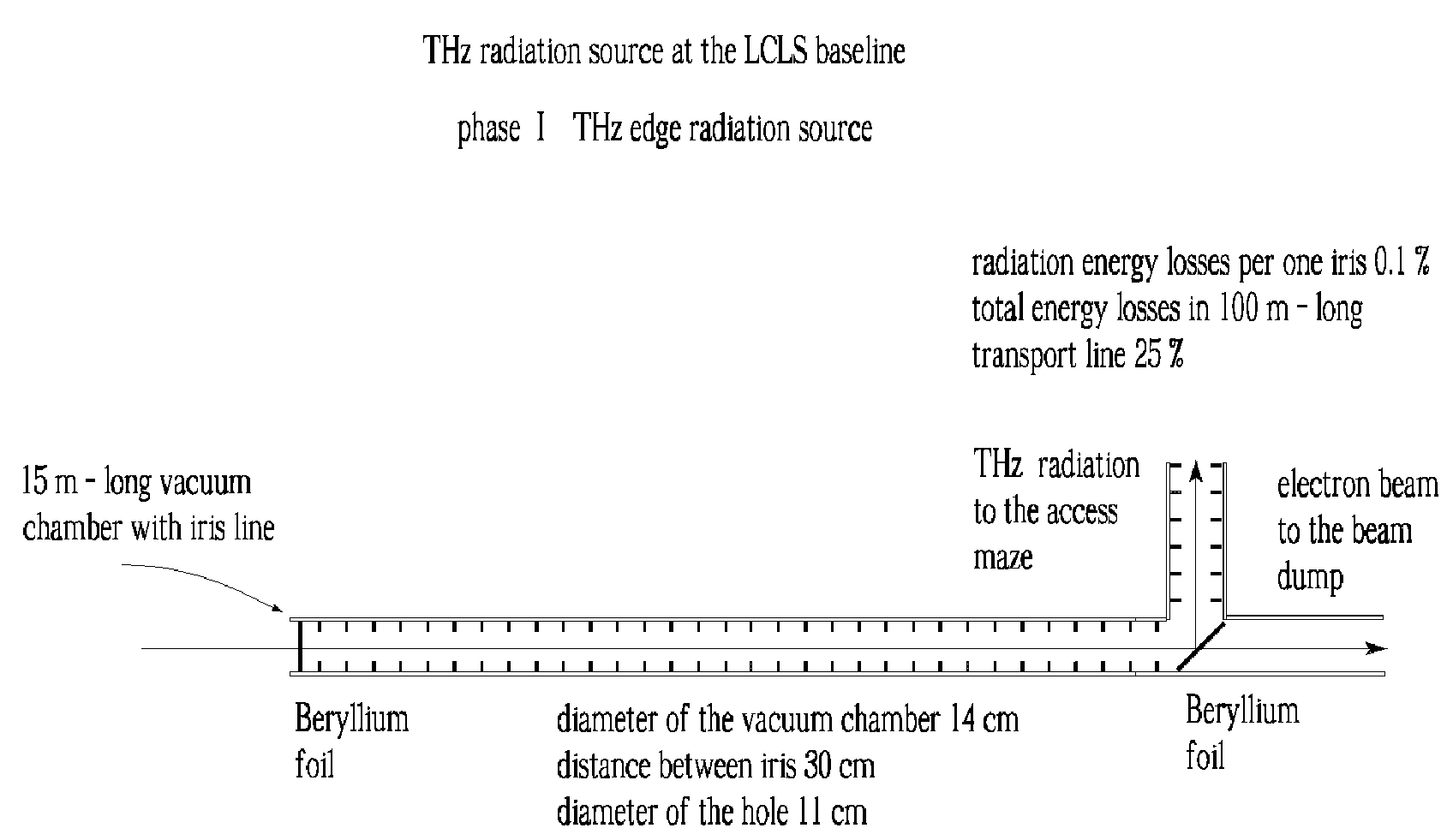}
\caption{Scheme of generation of edge radiation at the LCLS
baseline.} \label{th12}
\end{figure}

\begin{figure}[tb]
\includegraphics[width=1.0\textwidth]{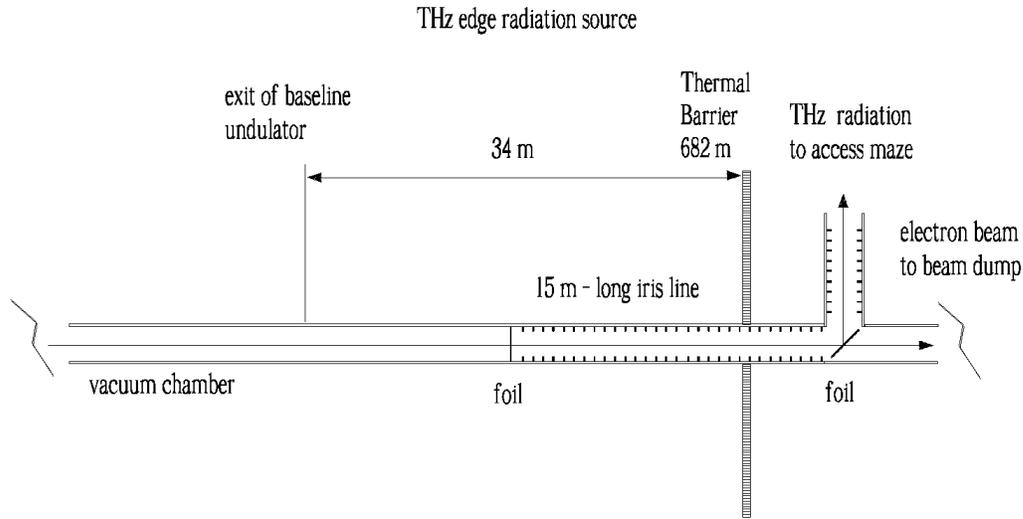}
\caption{Installation of the iris line for the THz edge radiation
source downstream of the LCLS baseline undulator.} \label{th13}
\end{figure}

\begin{figure}[tb]
\includegraphics[width=1.0\textwidth]{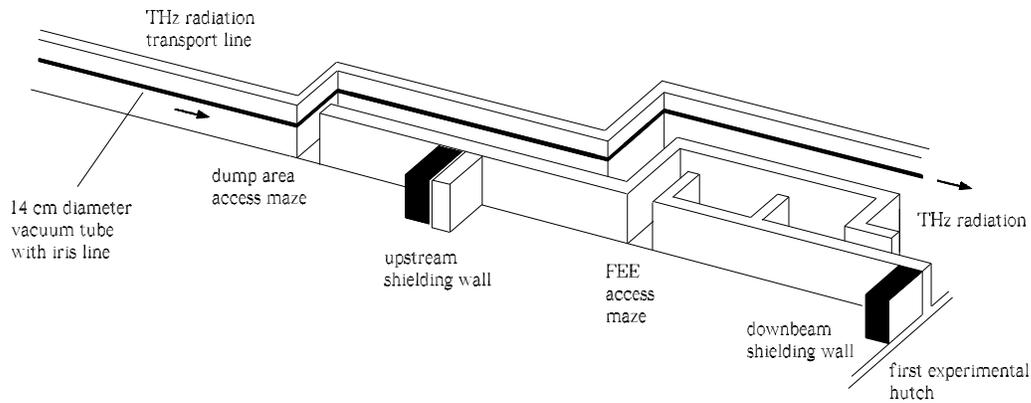}
\caption{Scheme of transport of the THz radiation to the
experimental X- ray floor.} \label{th14}
\end{figure}
The THz edge radiation source proposed in this paper is compatible
with the layout of the LCLS baseline and can be realized with
minimal additional efforts. The vacuum chamber equipped with iris
line and outcoupling system can be installed in the unoccupied
straight vacuum line behind the baseline undulator, Fig. \ref{th12}
and Fig. \ref{th13}. The transport of the THz beam in the LCLS
baseline case only be done with the use of the dump area access
maze, which is reproduced schematically in Fig. \ref{th14}.

\begin{figure}[tb]
\includegraphics[width=1.0\textwidth]{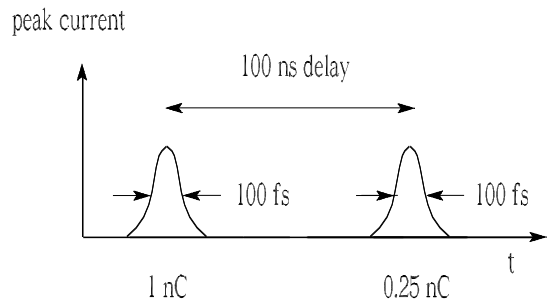}
\caption{Double electron bunch generation for THz pump/X-ray probe
experiments at the LCLS. The THz radiation transport through the
access maze introduces a path delay with respect to the X-ray path,
which has to be compensated with the introduction of a delay between
two bunches.} \label{th15}
\end{figure}
The THz transmission line transporting the THz radiation through the
access maze introduces a path-delay of about $30$ m. Since THz
pump/X-ray probe experiments should be enabled, we propose to
exploit a two-bunch mode of operation in order to cope with the
delay between THz and X-ray pulses, Fig. \ref{th15}. The two bunches
should be precisely separated by a certain temporal delay and such
delay should be matched to that introduced by the THz transmission
line. The main idea is, first, to produce two electron bunches at a
fixed distance with respect to each other. Second, the bunches are
accelerated and compressed. The beam energy and position is kept
nearly identical, while they progress one after the other through
the undulator and THz setup. A way to create two pulses delayed with
respect to each other is to split the UV laser pulse which
illuminates the cathode, delay one of the pulses, and then recombine
the pulses on the same trajectory heading towards the rf gun. One
method for accomplishing this is described in \cite{DING,GRIM}. A
multibunch mode of operation was also used at the SLAC linac
\cite{SEEM}. Routine operation with about $60$ ns spacing was
achieved. It should be noted that for the present scheme (in
contrast with that discussed in \cite{DING} ) we do not require an
essentially flat RF amplitude between bunches.

The operation of the proposed THz source is insensitive to the
emittance and energy spread of the electron beam. An analysis of the
parameters of the THz source shows that it will operate reliably
even for emittance and energy spread exceeding the LCLS
commissioning results by an order of magnitude.

\begin{figure}[tb]
\includegraphics[width=1.0\textwidth]{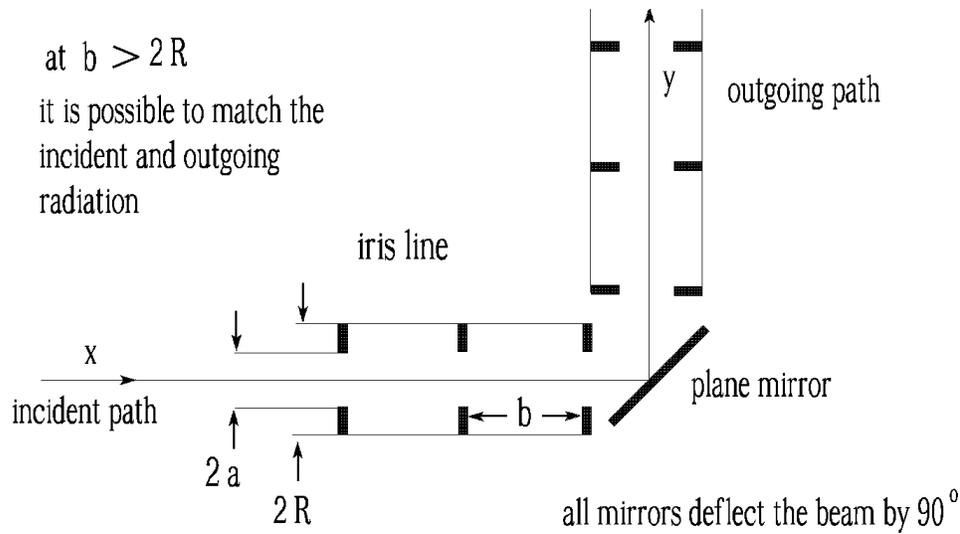}
\caption{Geometry of the transmission line turn. The losses are
calculated at the wavelength $\lambda = 0.1$ mm.} \label{th16}
\end{figure}
The THz transmission line for the LCLS baseline includes at least
six $90$ degrees turns, and will exploit plane mirrors as functional
components. If the pipe of the transmission line has a diameter
smaller than the distance between the irises  it is possible to
match incident and outgoing radiation without extra losses in these
irregularities, Fig. \ref{th16}.

\begin{figure}[tb]
\includegraphics[width=1.0\textwidth]{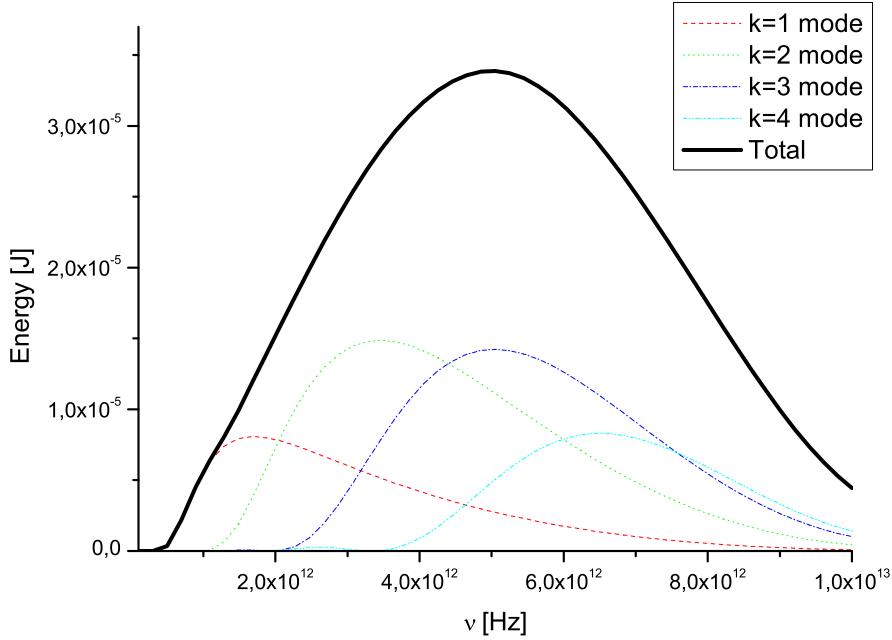}
\caption{Edge radiation pulse energy emitted within spectral
bandwidth $\Delta \omega/\omega = 10 \%$ and transported at the
sample position as a function of frequency. Partial contributions of
individual modes of the circular iris guide are illustrated. The
sample is set $100$ meters away from the extracting mirror. The
curves are calculated with analytical formulas in Eq. (\ref{loss})
and Eq. (\ref{energy}). Here $N_e = 6.4 \cdot 10^9$ ($1$ nC), $L =
15$ m, $b = 30$ cm, $2a = 11$ cm. The bunch form factor used is
shown in Fig. \ref{form}.} \label{ERpulse}
\end{figure}
It is possible to calculate the energy in the pulse, assuming a
$10\%$ spectral bandwidth. The bunch form factor considered here is
given in Fig. \ref{form}. We also consider a total length of the
iris line of $100$ m. The energy loss in the line are accounted for.
The energy per pulse at the exit of the iris line as a function of
the frequency $\nu = \omega/(2\pi)$ is shown in Fig. \ref{ERpulse}.

The maximal value of the pulse energy is achieved at  $\lambda
\simeq 0.06$ mm. When the bandpass filter is tuned  to  this value
of $\lambda$, the expression for the total edge radiation pulse
energy at the sample can be written in the form:

\begin{eqnarray}
W[\mathrm{mJ}] \simeq 0.34 \cdot \frac{\Delta \omega}{\omega}~.
\label{pulsesss}
\end{eqnarray}
The energy as a function of frequency exhibits a low frequency
cutoff due to losses in the transport line and a high frequency
cutoff due to form factor suppression.

\section{\label{sec:sette} Scheme for generating  THz undulator radiation from the LCLS baseline}

\begin{figure}[tb]
\includegraphics[width=1.0\textwidth]{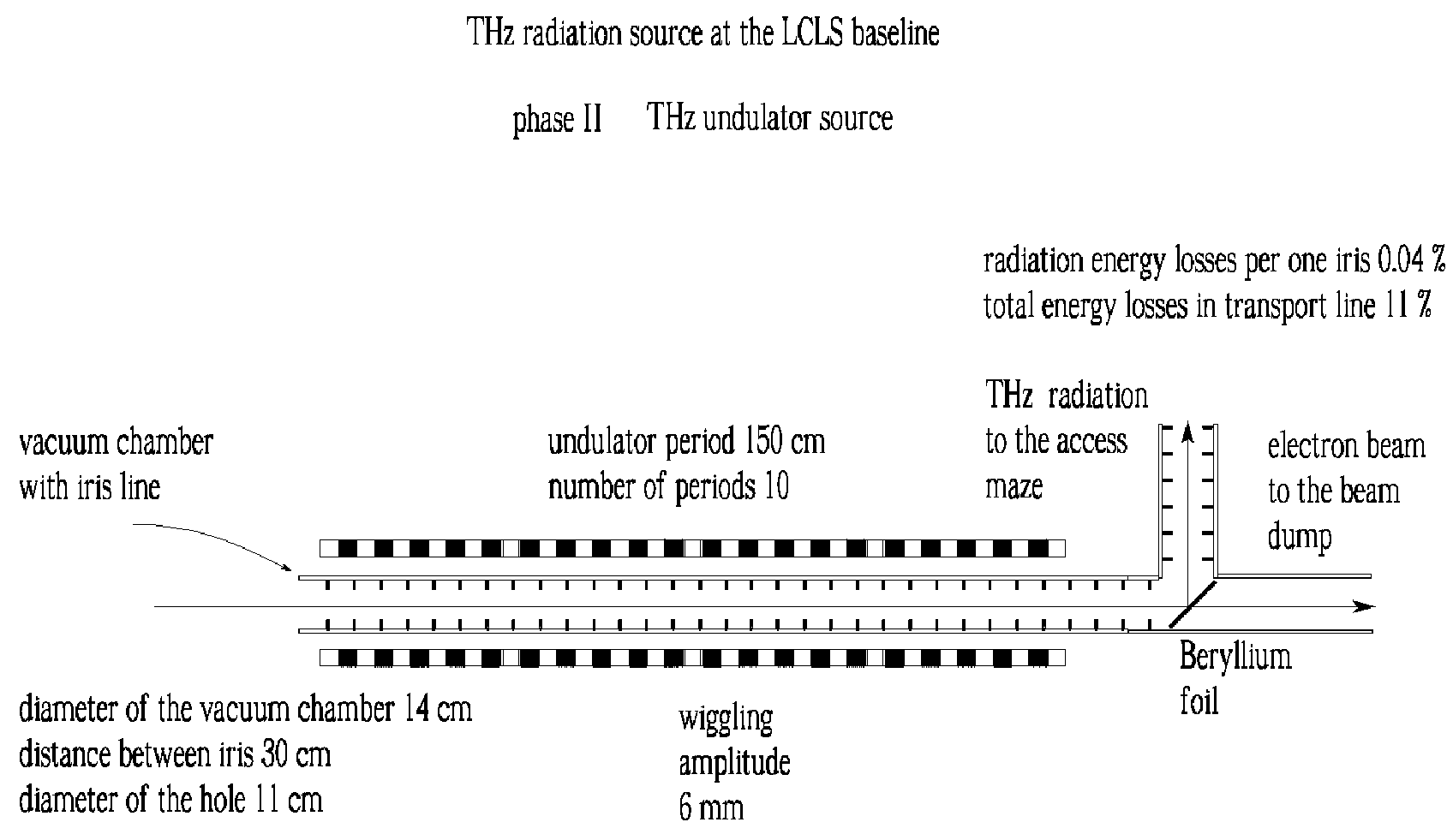}
\caption{Scheme for generating  THz undulator radiation at the LCLS
baseline. The installation of the THz undulator does not perturb the
edge radiation setup and does not interfere with the THz edge
radiation source operation.  The losses are calculated at the
wavelength $\lambda = 0.1$ mm.} \label{th18}
\end{figure}

\begin{figure}[tb]
\includegraphics[width=1.0\textwidth]{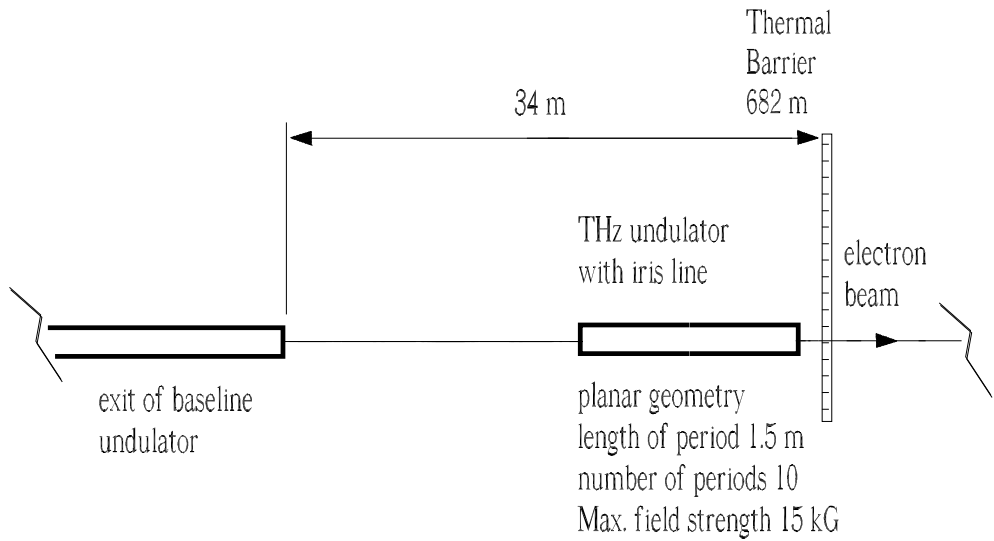}
\caption{Installation of the THz undulator downstream of the
baseline undulator.} \label{th17}
\end{figure}
As discussed in Section \ref{sec:due}, an optimal expansion strategy
for the LCLS THz source should be based on the installation of a THz
undulator, Fig. \ref{th18}, Fig. \ref{th17}. The undulator proposed
is an inexpensive planar electromagnetic device with $10$ periods,
each $1.5$ m long. At the electron beam energy around $5$ GeV and at
a THz wavelength around $0.2$ mm, the peak value of the magnetic
field is about $1.5$ T. Undulator radiation will be produced around
the fundamental harmonic, and the fundamental harmonic will be tuned
by adjusting the magnetic field strength. The wavelength range
between $0.05$ mm and $0.2$ mm provided by this powerful radiation
source will overlap with a large part of the THz gap. This will
allow applications for pump-probe experiments combining THz and
X-ray radiation\footnote{There is a trade-off between the
far-infrared and the X-ray frequency range achievable in the case a
THz undulator source is used. For instance, at an electron energy
around $10$ GeV the longest far-infrared and shortest X-ray
wavelengths available for operations are around $0.05$ mm and $0.25$
nm respectively.}.

Note that in the case of THz undulator source at LCLS the electron
beam emittance is much smaller than the radiation wavelength. This
means that the electron beam can safely be modeled as filament beam.
Computer codes like SRW \cite{SRW1} and SPECTRA \cite{SPEC} can be
used to study the characteristics of the output radiation in the
space-frequency domain up to a wavelength where the influence of the
vacuum chamber is negligible. However, in the case of the THz
undulator source at the LCLS, wavelengths of order of 0.1 mm and 15
m - long undulator yield a radiation diffraction size of order of 10
cm. This rough estimate indicates that vacuum chamber effects are
expected to play an important role. In this case, conventional
computer codes fail to predict the correct radiation
characteristics.

Summing up, in view of practical applications to the THz undulator
at LCLS, there is a need for a theory of undulator radiation in the
presence of a waveguide. In \cite{TUND} we developed a theory of
undulator radiation in the presence of the vacuum chamber with
circularly symmetric cross-section. In order to efficiently couple
radiation into the iris transmission line, for the LCLS case we
proposed to use a THz undulator source together with an iris line,
Fig. \ref{th18} and Fig. \ref{th17}.   The task that one has to
solve differs from the circular vacuum chamber case \cite{TUND} in
the formulation of boundary conditions. Using Leontovich boundary
conditions gives a good approximation in the case of a metallic
vacuum pipe with resistive walls, and drastically simplifies the
solution of the electrodynamical problem \cite{TUND}. The problem of
mode excitation in an iris waveguide can be solved following the
same approach, where Leontovich boundary conditions are now
substituted by Vainstein boundary conditions. In our practical case
of interest the wiggling amplitude of the electrons in the undulator
is small with respect to the dimension of the iris waveguide, Fig.
\ref{th17}. This greatly simplifies analytical calculations. In
free-space and under the resonance approximation undulator radiation
is horizontally polarized. This is actually a replica of the
undulator polarization properties. Moreover, the field exhibits
azimuthal symmetry. This properties  are the same when an iris line
is introduced \footnote{These properties are lost in the case of
metallic pipe when Leontovich boundary conditions are introduced
\cite{TUND}.}.

\begin{figure}[tb]
\includegraphics[width=1.0\textwidth]{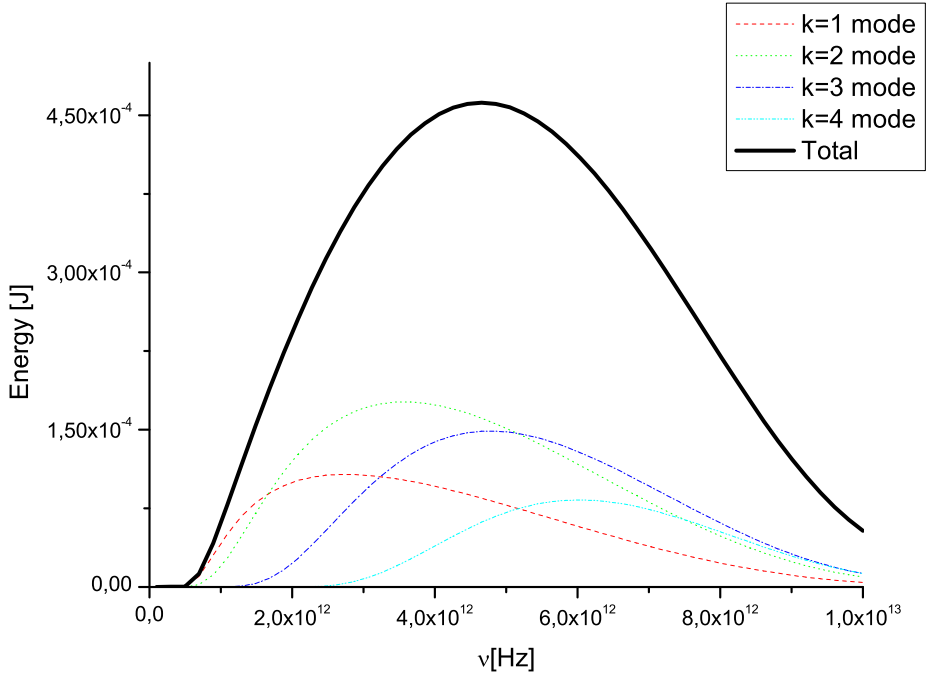}
\caption{Undulator radiation pulse energy emitted with a spectral
bandwidth $\Delta \omega/\omega = 10 \%$ and transported at the
sample position as  a function of frequency. Partial contributions
of individual modes of the circular iris guide are illustrated. The
sample is set $100$ meters away from the extracting mirror. The
curves are calculated with the analytical formulas in Eq.
(\ref{loss}) and Eq. (\ref{energyw}). Here $N_e = 6.4\cdot 10^9$
($1$ nC), $N_w = 10$, $\lambda_w = 1.5$ m, $b$ = 30 cm, $2a = 11$
cm. The bunch form factor used is shown in Fig. \ref{form}. Curves
correspond to the particular choice of the detuning parameter $C$ at
which the pulse energy in the second mode achieves its maximum.}
\label{URpulse}
\end{figure}
It is possible to calculate the energy in the pulse, assuming a
$10\%$ spectral bandwidth. The bunch form factor considered here is
given in Fig. \ref{form}. We also consider a total length of the
iris line of $100$ m. The energy loss in the line are also accounted
for. The energy per pulse at the exit of the iris line as a function
of the frequency $\nu = \omega/(2\pi)$ is shown in Fig.
\ref{URpulse}.

It can be easily shown that the energy per spectral interval, Eq.
(\ref{energyw}), does not depend on the energy of the electron beam
nor, in our case of interest with $K \gg 1$, on the undulator
parameter. In other words, we assume that the fundamental harmonic
will be tuned by adjusting the magnetic field strength at some fixed
electron beam energy, but there is no need to specify these
parameters.

It is instructive to calculate the ratio ${W}/{W_{\mathrm{fs}}}$ in
Eq. (\ref{finee}) and find the departure of the results presented in
Fig. \ref{URpulse} from the free space, ideal performance.
Calculating the Eq. (\ref{finee}) at the sample position and at the
optimal wavelength $\lambda \simeq 0.06$ mm, we obtain
${W}/{W_{\mathrm{fs}}} \simeq 0.8$ . This comparison demonstrates
the effectiveness of the proposed setup.

Finally, we note that according to our calculations, at a wavelength
around $0.06$ mm, and for ten cycles in the pulse,  one obtains a
0.3 GW peak power level on the sample (see Fig. \ref{URpulse}). With
strong focusing, these THz beams will approach the high field limit
of $1$ V$/$atomic size.

\section{\label{sedc:nove} Conclusions}

In this paper we presented a design of a THz edge radiation source
and a possible extension to a THz undulator radiation source for the
LCLS baseline. We began our considerations from the generation of
THz radiation from the spent electron beam downstream of the
baseline undulator. Transmission of the THz beam can only be
accomplished with quasi-optical techniques. In this article we
proposed to use an open beam waveguide such as an iris guide, and we
presented a complete theory of iris guides. In particular,
eigenmodes were studied both numerically and analytically. In order
to efficiently couple radiation into the iris transmission line, it
is desirable to match the spatial pattern of the source radiation to
the propagating mode of the transmission line. To solve the matching
problem, we proposed to generate the THz radiation within the iris
guide.

Following a derivation of the iris guide Green's function we
presented the the first exhaustive theory of edge and undulator
radiation within an iris guide, and we exemplified it in the case of
the proposed THz source at LCLS.  We specialized our consideration
to the case of a circular iris waveguide. The electric field was
found as a superposition of the iris waveguide modes, and was
studied for the LCLS parameters case.

The concepts presented here are applicable not only to the LCLS, but
also to other XFEL facilities. In fact, we regard the concepts
presented here, based on the combination of edge radiation and
undulator radiation with an open waveguide setup, as the start of a
novel direction towards the construction of THz sources at XFELs.
Therefore, the present work should not be considered as
comprehensive of all possible solutions. On the contrary, this work
is presented with the goal of stimulating interest and of opening
the door to new possibilities for THz radiation production and
transport at XFELs.

\section{Acknowledgements}

We are grateful to Massimo Altarelli, Reinhard Brinkmann, Serguei
Molodtsov and Edgar Weckert for their support and their interest
during the compilation of this work.

\end{document}